\documentclass[
  twocolumn,
  prd,
  reprint,
  nofootinbib,
  amsmath,
  amssymb,
  floatfix
]{revtex4-2}

\usepackage{url}
\usepackage{hyperref}
\hypersetup{
    colorlinks=true,
    linkcolor=blue,
    filecolor=magenta,      
    urlcolor=blue,
    pdftitle={Overleaf Example},
    pdfpagemode=FullScreen,
    }

\usepackage{enumitem}
\usepackage{graphicx}
\usepackage{dcolumn}
\usepackage{slashed}
\usepackage{float}
\usepackage{bm}
\usepackage{ulem}

\usepackage[caption=false]{subfig}

\def\beq{\begin{equation}}
\def\eeq{\end{equation}}
\def\be{\begin{equation}}
\def\ee{\end{equation}}
\def\bea{\begin{eqnarray}}
\def\eea{\end{eqnarray}}

\DeclareMathOperator{\mA}{\mathcal{A}}
\DeclareMathOperator{\mB}{\mathcal{B}}

\def\x{\chi}
\def\xd{\dot{\chi}}

\def\F{G}

\def\be{\bar{e}}

\newcommand{\kk}[1]{{\color{cyan}#1}}

\begin{document}
\title{3-form dark energy and cosmic birefringence}

\author{Kohei Kamada$^{1,2,3}$}
\email[Email: ]{ kohei.kamada$@$ucas.ac.cn}
\author{Tucker Manton$^{1}$}
\email[Email: ]{ tucker\_manton$@$ucas.ac.cn}
\affiliation{$^1$School of Fundamental Physics and Mathematical Sciences, Hangzhou Institute for Advanced Study,\\
\phantom{}\hspace{0.5cm} University of Chinese Academy of Sciences (HIAS-UCAS), Hangzhou, 310024, China.}
\affiliation{$^2$International Centre for Theoretical Physics Asia-Pacific (ICTP-AP),\\
\phantom{}\hspace{0.5cm}Hangzhou/Beijing, China.}
\affiliation{$^3$Research Center for the Early Universe, The University of Tokyo, 
Bunkyo-ku, Tokyo 113-0033, Japan.}

\date{\today}

\begin{abstract}
3-forms are interesting fields to study in the cosmological context for numerous reasons, such as being candidates for explaining inflation and dark energy. The background evolution of a 3-form field is similar to but distinguishable from a scalar field in an expanding universe, and its tensorial structure allows for unique couplings that cannot be experienced by canonical scalars. In this work, we explore the possibility that 3-form dark energy can explain cosmic birefringence. We consider two EFT-inspired couplings between the 3-form and the photon, and compute the birefringence angle $\beta$.  We find that a dimension-6 gauge-invariant operator necessitates an extremely large coupling to explain $\beta\sim0.3^\circ$, the value suggested by observations of recent cosmic microwave background radiation. Conversely, a dimension-4 operator can accommodate $\beta\sim0.3^\circ$ at the expense exciting the longitudinal mode of the photon. Interestingly, demanding a self-consistent decoupling limit implies the photon mass lies within a few orders of magnitude of $H_0$. We also derive  
`universal' profiles for $\beta(z)$, finding the $\beta$ from the dimension-4  operator is insensitive to the form of the 3-form potential but is sensitive to the initial conditions. Contrarily, the dimension-6 operator is highly sensitive to the form of the potential. We finally compare the universal profiles to that of  axion-like particle (ALP) dark energy and an ultralight massive 3-form, the latter obtained from numerically integrated cosmological histories consistent with $\Lambda$CDM up to low redshift. Our results show that birefringence from 3-form dark energy can both mimic that from an ALP or be distinguishable, depending on the field configuration in the early universe.
\end{abstract}

\pacs{Valid PACS appear here}
\maketitle
\section{Introduction}
The cosmic microwave background (CMB) carries an imprint of primordial physics not only in its temperature anisotropies, but also in its polarization patterns. A puzzling, large-scale signal 
observed in recent CMB data - interpreted as isotropic cosmic birefringence - suggests a rotation of CMB polarization planes between recombination and the present day, parameterized by an angle $\beta\approx0.3^\circ$ \cite{Komatsu:2022nvu}. More specifically, analysis of ACT DR6 finds $\beta=0.215^\circ\pm 0.074^\circ$ (68\% CL) \cite{Diego-Palazuelos:2025dmh}, while the combined analysis of Planck (PR4) + WMAP9 finds $\beta=0.342^\circ {}^{+0.094^\circ}_{-0.094^\circ}$ (68\% CL) \cite{Eskilt:2022cff} (see also \cite{Remazeilles:2025wzd}). The former excludes $\beta=0$ at 2.9$\sigma$ significance, the latter excluding at $3.6\sigma$. The combined analysis yields $\beta=0.264^\circ\pm0.058^\circ$, which excludes $\beta=0$ at $4.6\sigma$ significance. Although this result is very promising, the questions of systematic uncertainties, foreground contamination \cite{Clark:2021kze}, and polarization angle calibration \cite{Keating_2012} all remain. Additional, independent measurements are needed for cosmic birefringence to be a genuine discovery.

This effect provides a unique probe of parity-violating physics in the electromagnetic sector, acting between recombination and the present epoch. In particular, since the rotation angle $\beta$ accumulates over cosmological time, it encodes information about the time evolution of the underlying field responsible for the birefringence. The birefringence angle quantifies the mixing of $E-$ and $B-$modes, and can be observed by precision measurements of the $EB$-correlations in CMB polarization data. It is straightforward to relate $\beta$ to the dispersion relation $\omega_\pm(k)$ for the photon polarizations written in the helicity basis,
\begin{equation}\label{betadef}
\beta=-\frac{1}{2}\int_{t_{\text{LSS}}}^{t_0}\mathrm{d} t(\omega_+-\omega_-).
\end{equation}
Here, $t_{\text{LSS}}$ denotes the time of the surface of last scattering and $t_0$ denotes the present time. Any interaction that produces a nonzero $\beta$ must therefore violate parity; a parity preserving operator that modifies the photon dispersion relation will always contribute the same sign to $\omega_-$ and $\omega_+$. 

Crucially, the observed rotation angle does not show a frequency dependence 
\cite{Komatsu:2022nvu,Eskilt:2022cff,Eskilt:2022wav}. This effectively rules out Faraday rotation from magnetic fields from primordial physics \cite{Kosowsky:1996yc}, intergalactic medium \cite{Blasi:1999hu}, or galaxy clusters \cite{Carilli:2001hj} as being responsible for cosmic birefringence, which is a frequency \textit{dependent} phenomena. It is thus natural to attribute a nonzero rotation angle to the unknown physics of dark matter and/or dark energy. A well-known mechanism for generating cosmic birefringence is through an axion-like particle (ALP) coupled to the photon via a Chern-Simons interaction $\sim \phi F\tilde{F}$ \cite{Harari:1992ea,Carroll:1989vb,Lue:1998mq,Finelli:2008jv}. In this case, the rotation angle is proportional to the change in the axion field between recombination and today, $\beta \propto \Delta \phi$, a calculation we will review in Sec. \ref{sec:ALPbirefringence}. This simple structure has made axions a leading candidate for explaining the observed signal. 

Since any coupling to $F\tilde{F}$ produces cosmic birefringence, one can imagine that an Standard Model Effective Field Theory (SMEFT) operator may be able to be responsible for the phenomenon. In a very interesting work \cite{Nakai:2023zdr}, however, it was shown that a dimension-6 Higgs operator, dimension-7 lepton bilinear operator, or dimension-8 SM boson operators are all insufficient to produce $\beta\sim0.3^\circ$ under reasonable assumptions. The authors concluded that a new, light degree of freedom is necessary to explain cosmic birefringence. Although it is generally the case that a new (beyond Standard Model) degree of freedom is responsible for cosmic birefringence, alternatives such as Lorentz-violating electrodynamic terms \cite{Carroll:1989vb,Kostelecky:2007zz} and vacuum topology \cite{Kaloper:2026ygk} have also been proposed.

The analysis in \cite{Nakai:2023zdr} was restricted to only couplings to $F\tilde{F}$, and did not include the discussion of operators of the form $J_\mu A_\nu\tilde{F}^{\mu\nu}$ \cite{Geng:2007va,Ho:2010aq,Zhou:2023aqz}. Such operators generically induce a mass term for the photon, as we will discuss in depth in Sec. \ref{sec:dim4}, which is highly constrained by the Particle Data Group \cite{ParticleDataGroup:2024cfk}. Any interaction that breaks gauge invariance in the $U(1)_{\text{EM}}$ sector is of course physics beyond the Standard Model.

An additional puzzle in observational cosmology concerns the nature of late-time dark energy. Recent analysis of CMB+BAO is mildly suggestive of  a departure from $\Lambda$CDM  \cite{DES:2026jmi,DESI:2024lzq,DESI:2025zgx,eBOSS:2020yzd}. The deviations from $\Lambda$CDM are broadly referred to as dynamical dark energy (DDE), which would include an ultralight ALP. In this work, we explore whether or not cosmic birefringence and DDE can be caused by the same non-axionic degree of freedom, namely, 3-form dark energy. 

Cosmological 3-forms have been studied in the context of inflation and dark energy \cite{Koivisto:2009ew,Koivisto:2009fb,DeFelice:2012jt}, gravitational wave production \cite{Kobayashi:2009hj}, and quite recently, was confronted with observational data in \cite{Bouhmadi-Lopez:2025lzm,Bouhmadi-Lopez:2026ytj}. In that work, the authors studied a Gaussian potential, which results in phantom behavior. It was found that 3-form dark energy is a viable candidate for late-time acceleration and reducing the Hubble tension.

A 3-form in a homogeneous and isotropic background can be described by a single scalar degree of freedom we will call $\chi$ \cite{Heisenberg:2025jut}, however, its tensorial structure allows for interactions that are qualitatively distinct from those of scalar fields. In particular, the field strength of a 3-form naturally involves the combination $\dot{\chi} + 3H\chi$, which couples the field evolution directly to the expansion history of the universe. As a result, the birefringence signal generated by a 3-form can exhibit a different dependence on the cosmological background compared to the axion case. This makes 3-forms an interesting and well-motivated alternative for explaining the observed rotation angle.

We consider two effective interactions between the 3-form and the photon: a dimension-6 gauge-invariant operator and a dimension-4 operator that is not manifestly gauge invariant. We compute the resulting birefringence angle $\beta$ using realistic cosmological evolutions of the 3-form field. Our results show that the gauge-invariant dimension-6 operator requires an extremely large effective coupling to reproduce $\beta \sim 0.3^\circ$, placing it in tension with the expectations of effective field theory. In contrast, the dimension-4 interaction can naturally generate the observed signal, albeit at the cost of breaking $U(1)_{\rm EM}$ gauge invariance and exciting the longitudinal mode of the photon. This highlights a nontrivial interplay between symmetry principles and phenomenological viability in models of cosmic birefringence.

Importantly, the CMB is not the only probe of cosmic birefringence. If a birefringent medium permeates space, then any polarized cosmic light will exhibit a rotation angle as it propagates to earth. It is therefore possible to probe cosmic birefringence using radio galaxies at low redshifts \cite{Carroll:1989vb,Nodland:1997cc,Carroll:1997tc,Naokawa:2025shr} (see also Ref.~\cite{Yin:2024fez}). This simply amounts to modifying the integration bounds in \eqref{mainbeta} between some $t(z_i)$ to $t(z_f)$. The amount of birefringence of a given polarized source will in turn be sensitive to the shape of $\beta(z)$. ALP dark energy is known to produce a universal profile for $\beta(z)$, which is largely insensitive to the ALP mass or even potential shape, as long as it slow-rolls \cite{Naokawa:2025shr}. It is therefore of interest to compare the profiles for $\beta(z)$ from alternative explanations to ALP birefringence, which we address in Sec. \ref{sec:ProfilesAndUniversality}.

This note is organized as follows: in Sec. \ref{sec:ALPbirefringence}, we give a brief overview of axion birefringence. In Sec. \ref{sec:3formDE}, we introduce the 3-form field and overview its role as dark energy. We then develop an analytical understanding of two `universal' branches of field trajectories, large- and small-field, which are insensitive to the 3-form potential. We then introduce a set of bounded phase space variables   following \cite{Bouhmadi-Lopez:2016dzw}, and obtain specific field trajectories for the case of a massive potential. We here find an intermediate `middle-field' trajectory between the large- and small-field trajectories, and numerically solve the Einstein equations, observing cosmological histories consistent with $\Lambda$CDM up to low redshift. In Sec. \ref{sec:Interactions}, we work through the rotation angle calculation for two effective interactions between the 3-form and photon, and discuss their scaling properties. In Sec. \ref{sec:ProfilesAndUniversality}, we derive `universal' profiles for the 3-form $\beta(z)$, corresponding to the large- and small-field branches. For the universal profiles, the shape of $\beta(z)$ is also insensitive to the form of the potential. We then compare these profiles to the ALP dark energy case. Interestingly, the mid-field trajectories numerically integrated in Sec. \ref{sec:3formDE} produce shapes for $\beta(z)$ that are uniquely distinguishable from the ALP profile and large- and small-field profiles. 
We discuss and conclude in Sec. \ref{sec:Conclusion}.

We use a mostly plus metric signature $(-+++)$ and set $c=1$ throughout. 
\section{ALP birefringence}\label{sec:ALPbirefringence}
Originally proposed as an explanation of the strong CP problem \cite{Peccei:1977hh,Wilczek:1977pj,Weinberg:1977ma}, the axion was soon after shown to be a viable dark matter candidate \cite{Preskill:1982cy}. Axions are ubiquitous in string theory \cite{Arvanitaki:2009fg,Svrcek:2006yi} and have been studied as composite particles \cite{Kim:1984pt,Alexander:2023wgk,Alexander:2024nvi}. The many different types of axions are collectively referred to as  axion-like particles (ALPs), and numerous ongoing experiments are searching for direct and indirect evidence of ALPs (for a review, see \cite{Marsh:2015xka}).

An ALP $\phi$ is described by the Lagrangian $\mathcal{L}_\phi=-\tfrac{1}{2}(\nabla_\mu\phi)^2-V(\phi)+\mathcal{L}_{int}$. 
The potential $V(\phi)$ is typically taken to be of a cosine form $M^4 (1-\cos(\phi/f_\phi))$ or a simple mass term $m^2_\phi \phi^2/2$, the lowest order of expansion. The common trait shared by all ALPs is $\mathcal{L}_{int}$ includes the Chern-Simons interaction with the photon, $\mathcal{L}_{int}=-\tfrac{\alpha}{4f_\phi}\phi F\tilde{F}$. 
$F_{\mu\nu}$ is the electromagnetic field strength tensor and 
$\tilde{F}^{\mu\nu}=\tfrac{1}{2\sqrt{-g}}\epsilon^{\mu\nu\rho\sigma}F_{\rho\sigma}$ is its dual. Here, $g$ denotes the determinant of the metric, and $\epsilon^{\mu\nu\rho\sigma}$ is the totally anti-symmetric tensor with $\epsilon^{0123} = 1$.
$f_\phi$ is the axion decay constant, and $\alpha$ is a dimensionless coupling constant related to
the fine structure constant but also dependent on the underlying UV physics.
Neglecting the backreaction from $\mathcal{L}_{int}$, the  equation of motion for an homogeneous axion in a cosmological background is
\begin{equation}
    \ddot{\phi}+3H(t)\dot{\phi}+\frac{\mathrm{d}V}{\mathrm{d} 
    \phi}=0, 
\end{equation}
where $H(t)$ is the Hubble parameter.

The photon field is decomposed into Fourier modes as
\begin{equation}\label{photonfield}
    A_i^\pm(x,t)=\sum_{s=\pm}\int\frac{\mathrm{d}^3k}{(2\pi)^3}A_{k,s}(t)\varepsilon^s_i e^{\mathrm{i}kx},
\end{equation}
where $\vec{\varepsilon}_\pm$ are the polarization vectors in the circular basis satisfying $\vec{\varepsilon}_\pm\cdot\vec{k}=0$, $\vec{\varepsilon}_\pm\cdot\vec{\varepsilon}_\mp=0$ and importantly, $\vec{k}\times\vec{\varepsilon}_\pm=\pm\tfrac{k}{\mathrm{i}}\vec{\varepsilon}_\pm$ with $k \equiv |\vec{k}|$; the cross-product between the photon momentum and polarization at the level of the equations of motion contributes the opposite sign to the dynamics of the fluctuations $A_{k,\pm}(t)$, 
leading to cosmic birefringence. 

The ALP--photon interaction results in the photon field equation of motion on top of the background ALP dynamics (in cosmic time $t$),
\begin{equation}
    \ddot{A}_{k,\pm}+H\dot{A}_{k,\pm}+\Big(\frac{k^2}{a^2}\mp\frac{k}{a}\frac{\alpha\dot{\phi}}{f_\phi}\Big)A_{k,\pm}=0,
\end{equation}
where we have adopted the radiation gauge, $A_0 = \partial^i A_i=0$.
The corresponding dispersion relations are $\omega_\pm=\frac{k}{a}\mp\frac{\alpha}{2f_\phi}\dot{\phi}$ in the WKB approximation. The two circular polarization modes propagate with different phase velocities, leading to the rotation of the polarization angle and cosmic birefringence. 
This produces the well known result for the total rotation angle \eqref{betadef},
\begin{equation}\label{axionbeta}
    \beta = \frac{\alpha}{2f_\phi}\int_{t_\mathrm{LSS}}^{t_0}\mathrm{d}t\dot{\phi}=\frac{\alpha}{2f_\phi}(\phi(t_0)-\phi(t_\mathrm{LSS})), 
\end{equation}
where $t_\mathrm{LSS}$ and $t_0$ is the cosmic time at the last scattering surface and today, respectively.
Famously, the rotation angle stemming from an ALP interaction depends only on the difference in the field values at the endpoints. Note that if the field value changes largely between reionization and recombination, the shape of the CMB EB power spectra is modified significantly such that it provides the information of the ALP dynamics~\cite{Finelli:2008jv,Sherwin:2021vgb,Nakatsuka:2022epj}, 
which is the case of ALP dark matter.
In the case ALP dark energy, on the other hand, the CMB spectra is relatively insensitive to the ALP dynamics.

The CMB is not the unique probe of the cosmic birefringence\footnote{Indeed, a very recent proposal \cite{Matsuoka:2026qya} suggested observing nearby binary systems to probe ALP dark matter.}.
It has been known that radio galaxies with strong jets emit linearly polarized photons, which can be used to detect the cosmic birefringence in a tomographic way at $z\lesssim 10$~\cite{Carroll:1989vb,Nodland:1997cc,Carroll:1997tc,Naokawa:2025shr}. 
If the ALP is slow-rolling from the last scattering surface until today,
as is typical in the case where it is responsible for dark energy, 
the information of the ALP dynamics can be obtained from this probe. Remarkably, the redshift-dependent rotation angle, $\beta(z) = (\alpha/2 f_\phi) \int_{t(z)}^{t_0} \mathrm{d} t {\dot \phi}$, shows a universal profile~\cite{Naokawa:2025shr},
\begin{equation}
    \beta(z) \sim \beta (z_\mathrm{LSS})  \int_0^z \frac{\mathrm{d}z}{(1+z)^4 ( 1 - \Omega_{\Lambda 0})+(1+z) \Omega_{\Lambda 0}}, \label{ALPUnivProfile}
\end{equation}
where $\Omega_{\Lambda_0}$ is the present energy density fraction of the cosmological constant. 
A significant deviation from this profile would indicate that the reported cosmic birefringence is generated by a mechanism other than a slow-rolling ALP, including one responsible for dark energy. 
As we will see, our 3-form interactions can significantly depart from this picture, leading to qualitatively different birefringence signatures.

\section{3-form cosmology}\label{sec:3formDE}
We will denote our 3-form field as $C_{\mu\nu\rho}$. It is described by the Lagrangian
\begin{equation}
    \mathcal{L}_C=-\frac{1}{48}\F^{\mu\nu\rho\sigma}\F_{\mu\nu\rho\sigma}-U(C^2),
\end{equation}
where $C^2=C^{\mu\nu\rho}C_{\mu\nu\rho}$, 
and the field strength $\F=dC$ has components
\begin{equation}
    \F_{\mu\nu\rho\sigma}=\nabla_\mu C_{\nu\rho\sigma}-\nabla_\sigma C_{\mu\nu\rho}+\nabla_\rho C_{\sigma\mu\nu}-\nabla_\nu C_{\rho\sigma\mu}.
\end{equation}
The potential $U(C^2)$ explicitly breaks gauge symmetry $C_{\mu\nu\rho} \rightarrow C_{\mu\nu\rho} +\partial_{[\mu} \Lambda_{\nu\rho]}$, which itself possesses the two-fold redundancies $\Lambda_{\mu\nu}\rightarrow \Lambda_{\mu\nu}+\partial_{[\mu}B_{\nu]},$ and $B_\mu\rightarrow B_\mu+\partial_\mu\lambda.$ Breaking the gauge symmetry by the introduction of a potential is essential for the 3-form's behavior to differ from a cosmological constant, as we will see below.   

Note that the this theory is dual to a theory of a vector $V^\mu\sim\epsilon^{\mu\nu\rho\sigma}C_{\nu\rho\sigma}$, with Lagrangian $\mathcal{L}_V=-\frac{1}{2}(\nabla_\mu V^\mu)^2-U(V^2)$. For our purposes, it is more convenient to work with the tensor $C_{\mu\nu\rho}$.

The free theory is given by
\begin{equation}
    S=\int \mathrm{d}^4x\sqrt{-g}\Big(\tfrac{M_p^2}{2}R+\mathcal{L}_C+\mathcal{L}_m+\mathcal{L}_r\Big),
\end{equation}
where $M_p$ is the reduced Planck mass, and $\mathcal{L}_m$ and $\mathcal{L}_r$ encompass the additional matter and radiation content, respectively.  
The Einstein equations are
\begin{equation}
M_p^2 G_{\mu\nu}
=
T_{\mu\nu}^{(C)} + T_{\mu\nu}^{(m)} + T_{\mu\nu}^{(r)},
\end{equation}
where the energy-momentum tensor of the three-form is
\begin{align}
T_{\mu\nu}^{(C)} ={}&
\frac{1}{6} G_{\mu\alpha\beta\gamma} G_{\nu}^{\ \alpha\beta\gamma}
+ 2\,U_{,C^2}\, C_{\mu\alpha\beta} C_{\nu}^{\ \alpha\beta} \\
&- g_{\mu\nu}\Big[\frac{1}{48} G^{\alpha\beta\gamma\delta} G_{\alpha\beta\gamma\delta}
 + U(C^2)\Big],
\end{align}
with $U_{,C^2} \equiv dU/d(C^2)$. Varying the action with respect to $C_{\mu\nu\rho}$ gives the covariant equation of motion
\begin{equation}\label{Ceom}
\nabla_\sigma G^{\sigma\mu\nu\rho}
=
12\,U_{,C^2}\, C^{\mu\nu\rho}.
\end{equation}

We will restrict our attention to the case where the 3-form field's evolution is dominated by its background equations. In a homogeneous, isotropic spacetime, the 3-form field can be described by a single, scalar degree of freedom depending only on time, $\chi(t)$ \cite{Heisenberg:2025jut}. In a flat, FLRW cosmological background with scale factor $a(t)$, the spatial components $C_{ijk}$ of the 3-form field have fluctuations parameterized by
\begin{equation}\label{Ccomps}
    C_{ijk}=a^3(t)\chi(t)\epsilon_{ijk}, \ \ \ \ C_{0ij}=0.
\end{equation}
The invariant $C^2=6\chi^2,$ so we can equivalently write the potential as $U(\chi^2)$. The nonzero components of the field strength are 
\begin{equation}\label{Gcomps}
    G_{0ijk}=a^3(t)[\dot{\chi}(t)+3H(t)\chi(t)]\epsilon_{ijk},
\end{equation}
where $H=\dot{a}/a$ is the Hubble parameter. The free field equation for $\chi$ is found by inserting \eqref{Ccomps} into \eqref{Ceom}, obtaining \cite{Bouhmadi-Lopez:2025lzm,DeFelice:2012jt,Koivisto:2009ew,Koivisto:2009fb}
\begin{equation}\label{chiEOM}
    \ddot{\chi}(t)+3\frac{\mathrm{d}}{\mathrm{d}t}[H(t)\chi(t)]+\frac{\mathrm{d}U}{\mathrm{d}\chi}=0.
\end{equation}
A critical difference between a canonical scalar field and the 3-form scalar is the presence of $\dot{H}$ in the equation of motion. 

Following from the Einstein equations, the energy density and pressure of the 3-form field are
\begin{equation}
    \begin{split}
        \rho_\chi = -T^0_0&=\frac{1}{2}(\dot{\chi}+3H\chi)^2+U(\x), \\
        p_\chi=\frac{1}{3}T^i_i&=-\frac{1}{2}(\xd+3H\x)^2-U(\x)+\x U'(\x),
    \end{split}
\end{equation}
so that the Friedmann equations are
\begin{equation}\label{firstFriedmann}
    H^2=\frac{1}{3M_p^2}\Big(\rho_m+\rho_r+\frac{1}{2}(\xd+3H\x)^2+U(\x)\Big),
\end{equation}
\begin{equation}\label{secondFriedmann}
    \dot{H}=-\frac{1}{2M_p^2}\Big(\rho_m+\frac{4}{3}\rho_r+\x U'(\x)\Big),
\end{equation}
where $\rho_m,\rho_r$ are the energy densities of matter and radiation.

We can use \eqref{secondFriedmann} to eliminate $\dot{H}$ in \eqref{chiEOM} to write the equation of motion for the 3-form scalar as
\begin{equation}
    \ddot{\x}+3H\dot{\x}+\Big(1-(\x/\x_c)^2\Big)\frac{\mathrm{d}U}{\mathrm{d}\x}=\frac{3}{2M_p^2}\Big(\rho_m+\frac{4}{3}\rho_r\Big)\chi, \label{chiEOM2}
\end{equation}
where $\x_c\equiv\sqrt{\frac{2}{3}}M_p$. After matter and radiation are predominantly redshifted away, the free equation for $\x$ has two critical points. The first corresponds to $U'=0$, as is standard for a canonical scalar field, while the second corresponds to $\x=\x_c=\sqrt{\frac{2}{3}}M_p$. In performing a dynamical phase space analysis, it is therefore natural to write the field as $\x=\x_cf(v)$ \cite{Bouhmadi-Lopez:2025lzm,Bouhmadi-Lopez:2016dzw, Koivisto:2009fb}, where $f$ is a function of a bounded, dimensionless phase space variable $v$. We similarly introduce for the Hubble parameter $H=H_0\tilde{f}(h)$, where $\tilde{f}$ is a function of a second, dimensionless phase space variable $h$. Using the dynamical phase space language is particularly useful in understanding the scaling behavior of our rotation angles, since $f=\x/\x_c$ and $\tilde{f}=H/H_0$ involve the IR and UV gravitational scales $H_0$ and $M_p$. In practice, the dynamical systems analysis is carried out in the dimensionless e-folding number $N=\log(a/a_0)=-\log(1+z).$ We will see examples of this in Sec. \ref{sec:Interactions}, as well as the explicit form of $f(v)$ and $\tilde{f}(h)$ in Sec. \ref{sec:phasespace}.

The sound speed is given by $c_s^2=\frac{\x U''(\x)}{U'(\x)},$ and the equation of state from the 3-form is 
\begin{equation}
    w_\x=\frac{p_\x}{\rho_\x}=-1+\frac{\x U'(\x)}{\tfrac{1}{2}(\xd+3H\x)^2+U(\x)}. \label{3formEOS}
\end{equation}
Depending on the nature of the potential, $w_\x$ can be greater than or less than $-1.$ In particular, whenever $U'(\chi)\approx0$, the 3-form field mimics a cosmological constant with $w_\x =-1$. This situation is similar to the case of a flat potential in canonical, slow roll inflation corresponding to (quasi) de Sitter expansion. 

\subsection{Analytic description of cosmological evolution of 3-form field} \label{subsec:analdyn3form}

Let us now investigate the dynamics of the 3-form field acting as dark energy, 
which is also based on the discussion in Ref.~\cite{Koivisto:2009ew}. 
As we have seen, $\chi=\chi_c$ is a critical point (see Eq.~\eqref{chiEOM2}).  
This suggests that a sufficiently long accelerating expansion of the Universe 
can be realized around $\chi\simeq \chi_c$ 
even for a relatively large $\mathrm{d}^2 U/\mathrm{d} \chi^2$, which is of our interest. 
On the other hand, during matter and radiation dominated eras, the right-hand side 
of Eq.~\eqref{chiEOM2}, which acts as a tachyonic mass,  
inevitably dominates the potential term for small $\mathrm{d} U/\mathrm{d} \chi$
and determines the 3-form field dynamics.  

Since we are interested in cosmic birefringence indicated by CMB measurements, 
we shall investigate the 3-form field dynamics after recombination. 
Approximating that the Universe is matter dominated and omitting the potential term,
the equation of motion is rewritten as
\begin{equation}
    \frac{\mathrm{d}}{\mathrm{d}N} \left[ \frac{\mathrm{d}\chi}{\mathrm{d} N} + 3 \chi \right] - \frac{3}{2} \left[ \frac{\mathrm{d}\chi}{\mathrm{d} N} + 3 \chi \right] =0, 
\end{equation}
where $N \equiv \int H  \mathrm{d}t = \log (a_0/a)$ is the number of $e$-folds.
This equation has two solutions that approaches to $\chi = \chi_c$ at the dark matter - 
dark energy equality, 
that is, a time-decreasing solution, 
\begin{equation}
    \chi_{\mathrm{lf}} = \chi_\mathrm{eq}^{\mathrm{lf}} \exp[- 3 (N-N_\mathrm{eq})] \propto H^2, \quad \text{with} \quad \chi_\mathrm{eq}^{\mathrm{lf}} \gg \chi_c, \label{eq:lfanal}
\end{equation}
and  a time-increasing solution
\begin{equation}
    \chi_{\mathrm{sf}} = \chi_\mathrm{eq}^{\mathrm{sf}} \exp[3 (N-N_\mathrm{eq})/2] \propto H^{-1}, \quad \text{with} \quad \chi_\mathrm{eq}^{\mathrm{sf}} \ll \chi_c, \label{eq:sfanal}
\end{equation}
which we call the large-field and small-field solutions, respectively. 
Here the subscript eq denotes that the variable is evaluated at the matter-radiation equality. 
In the expanding universe, the small-field solution can be thought of as an attractor in the sense that the large-field solution ultimately converges onto the small-field trajectory; the small-field solution at the recombination can have experienced the large-field dynamics at an earlier time. 
We also have the solution that is time-decreasing at the matter-radiation equality and turns around before the dark energy domination 
such that it becomes time-increasing at the dark energy - matter equality. This `middle-field' case can be understood as a superposition of the small and large field cases,
\begin{equation}\label{midfieldsolution}
    \chi_{\text{mf}}\simeq\mA\x_{\text{sf}}+\mB\x_{\text{lf}},
\end{equation}
with $\mathcal{A}$ and $\mathcal{B}$ determined by the initial conditions.
Their ratio also sets the time at which point the middle-field solution converges onto the small-field solution. This transition occurs at $e$-folding number $N_*$, when $\x'(N_*)\simeq0.$ Assuming $\mathcal{A}$ and $\mathcal{B}$ are positive, we can write this in terms of redshift as 
\begin{equation}\label{ABratio}
\frac{1+z_*}{1+z_{\text{eq}}}\simeq\Bigg(\frac{\mA}{2\mB}\Bigg)^{2/9}.
\end{equation}
We will see explicit examples of the cosmic birefringence angle stemming from each of these cases in Sec. \ref{sec:ProfilesAndUniversality}. Interestingly, the shape of $\beta(z)$ is significantly different for each field trajectory, and therefore is sensitive to the coefficients $\mA$ and $\mB$.

Note that we require that $\chi \simeq \chi_c$ at the dark energy - matter equality, 
we find
\begin{equation}
    \chi_\mathrm{eq}^{\mathrm{lf}} \simeq \chi_c \exp[3(N_\mathrm{Deq} - N_\mathrm{eq})], 
\end{equation}
for the large-field solution, and
\begin{equation}
    \chi_\mathrm{eq}^{\mathrm{sf}} \simeq \chi_c \exp[-3(N_\mathrm{Deq} - N_\mathrm{eq})/2], 
\end{equation}
for the small-field solution, 
where the subscript Deq means that the variable is evaluated at the dark energy - matter equality. 
We also note that 
\begin{equation}
    \frac{\mathrm{d}\chi}{\mathrm{d} N} + 3 \chi \simeq 0, 
\end{equation}
for the large-field solution, 
and 
\begin{equation}
    \frac{\mathrm{d}\chi}{\mathrm{d} N} + 3 \chi  = \frac{9\chi}{2} = \frac{9}{2} \chi_\mathrm{eq}^{\mathrm{lf}} \exp[3 (N-N_\mathrm{eq})/2], 
\end{equation} 
or 
\begin{equation}
    {\dot \chi} + 3 H \chi = \frac{9}{2}  H_\mathrm{eq} \chi_\mathrm{eq}^{\mathrm{sf}} = \mathrm{const.}, 
\end{equation}
for the small-field solution. 
In the former case, 
3-form field has a relatively large velocity, ${\dot \chi} \simeq - 3 H \chi$. 
In the latter case, at the time of the dark energy - matter equality 
the ``kinetic'' energy is comparable to the total energy density of the Universe,
$({\dot \chi} + 3 H \chi)^2/2 \simeq 3 H^2 M_p^2$, 
and hence the potential energy is negligible at that time.  

After the dark energy domination, 
it causes quasi de Sitter expansion of the Universe 
if $|\chi U^\prime(\chi)| \ll \rho_\chi \sim 3 H^2 M_p^2$. 
The matter or radiation contribution in the 3-form equation of motion 
is now negligible,
\begin{equation}
    \ddot{\x}+3H\dot{\x}+\Big(1-(\x/\x_c)^2\Big)\frac{dU}{d\x}=0. 
\end{equation}
We would expect that eventually the dynamics of the 3-form field is dominated by 
the potential term such that it is attracted to the critical point $\chi=\chi_c$
with slow-roll dynamics~\cite{Bouhmadi-Lopez:2025lzm,Koivisto:2009ew,Koivisto:2009fb,DeFelice:2012jt}. 
However, in both small and large-field solutions, ${\dot \chi}$ is relatively 
large at the time of dark energy - matter equality, and hence 
the dynamics is still well approximated by those of the matter-dominated era.

An interesting feature of the 3-form dark energy is that it allows for a relatively large deviation
from $\Lambda$CDM, $w=-1$, 
and a time-varying equation of state, depending on the potential, see Eq.~\eqref{3formEOS}. 
If the potential slope is not so small, 
$|\chi U^\prime(\chi)|/3 H_0^2 M_p^2 \sim \mathcal{O}(0.1)$, 
cosmological Large-Scale surveys can detect it as the deviation from the $\Lambda$CDM.
Interestingly, if $\chi U^\prime(\chi)$ is negative, 
it can cause a phantom behavior~\cite{Morais:2016bev,Bouhmadi-Lopez:2025lzm}, which is of interest in light of the 
recent reports from DES and DESI~\cite{DES:2026jmi,DESI:2024lzq,DESI:2025zgx}, 
with, however, a potential risk of the appearance of a ghost-like degree of freedom~\cite{DeFelice:2012jt}. 



\subsection{Phase space analysis}\label{sec:phasespace}

Here, we are interested in the time evolution of $\x$ and $H$ for a given choice of potential. To obtain them numerically, we follow the dynamical system analysis from \cite{Bouhmadi-Lopez:2016dzw}. There, the authors introduce five phase space variables $\{v,y,h,r,s\}$, which are related to $\x,\ {\dot \chi}, \ \rho_m, \ \rho_r$, and $H$ as
\begin{equation}\label{phasespacevars}
    \begin{split}
        &\hspace{15mm} x\equiv\frac{\x}{\x_c}=\frac{v}{1-v^2}, \ \ \ \ \ y=\frac{\dot{\x}+3H\x}{3H\x_c}, \\ 
        & h=\frac{(H/H_0)^2}{1+(H/H_0)^2}, 
         \ \ \ \ \ \    r=\frac{\sqrt{\rho_r}}{\sqrt{3}HM_p}, \ \ \ \ \ \ s=\frac{\sqrt{\rho_m}}{\sqrt{3}HM_p}.
    \end{split}
\end{equation}
These variables are functions of the $e$-folding number $N,$ and are bounded as $-1\leq \{v,y\}\leq 1$ and $0\leq \{h,r,s\}\leq 1.$ They must satisfy the Friedmann constraint, which can be written
\begin{equation}\label{friedmannconstraint}
    y^2+r^2+s^2+\frac{1-h}{h}\mathcal{U}(x)=1,
\end{equation}
where we have defined the dimensionless potential 
\begin{equation}\label{dimlesspotential}
    \mathcal{U}(x)\equiv\frac{1}{3M_p^2H_0^2}U(\x).
\end{equation}
The dynamics are then described by the system of equations 
\begin{equation}
    \begin{split}
        v'&=3\frac{1-v^2}{1+v^2}[y(1-v^2)-v], \\
        y'&=\frac{3}{2}\Big\{y\big(1-y^2+\frac{r^2}{3}\big)-\frac{1-h}{h}\big(\mathcal Uy+\mathcal U_x
\frac{1-v^2-vy}{1-v^2}\big)\Big\}, \\
h'&=-3(1-h)\big\{h\big(1-y^2+\frac{r^2}{3}\big)-(1-h)\big(\mathcal U-x\mathcal U_x\big)\big\}, \\
r'&=\frac{3}{2}r\Big\{-\frac{1}{3}-y^2+\frac{r^2}{3}-\frac{1-h}{h}\big[\mathcal U-x\mathcal U_x\big]\Big\},
    \end{split}
\end{equation}
with the variable $s$ determined by the Friedmann constraint \eqref{friedmannconstraint}. Here, the derivatives are with respect to the $e$-folding number, and 
$$\mathcal{U}_x=\frac{d\mathcal{U}}{dx}\Big|_{x=v/(1-v^2)}.$$

As discussed in the previous subsection, there are two primary branches of cosmological evolution, large- and small-field, with an intermediate case that lies in the middle. 
Since our interest is in cosmic birefringence, we will present three representative trajectories that are consistent with $\Lambda$CDM instead of doing an exhaustive phase space analysis. For convenience, we set initial conditions at matter-radiation equality, as was done in \cite{Bouhmadi-Lopez:2025lzm} for the case of a Gaussian potential. To ensure we begin in a region of phase space that is consistent with $\Lambda$CDM until low redshift, we constrain the energy density of the 3-form field to be the same order as $\Omega_\Lambda(z_\mathrm{eq})$. It is straightforward to show that the fractional energy density $\Omega_\chi=\rho_\x/3M_p^2H^2$ can be written in the simple form,
\begin{equation}
    \Omega_\x(y,h,x)=y^2+\frac{1-h}{h}\mathcal{U}(x).
\end{equation}
Using the Planck 2018 values $\Omega_{m0}\simeq0.311$ and $\Omega_{\Lambda 0}\simeq 0.689$ \cite{Planck:2018vyg}, and that $\rho_r\simeq\rho_m$ at $z_\mathrm{eq}\sim 3400,$ we have  $\Omega_{\chi,\mathrm{eq}}\sim O(10^{-11}).$ 

We here will focus on the case of a simple mass term,
\begin{equation}\label{massivepotentials}
    U(\x)=\frac{1}{2}m_\x^2\x^2, \ \ \ \ \ \mathcal{U}(x) = M^2x^2,
\end{equation}
with $M^2\equiv m_\x^2\x_c^2/6M_p^2H_0^2=m_\x^2/9H_0^2.$ The energy density constraint for the initial conditions $\{y_i,h_i,x_i\}$ is therefore
\begin{equation}\label{Omegainitial}
    y_i^2+\frac{1-h_i}{h_i}M^2x_i^2\simeq O(10^{-11}). 
\end{equation}
This condition simply restricts the balance of kinetic energy from $y$, potential energy $M^2x^2$, and the initial ratio of Hubble scales, since $(1-h)/h=H_0/H.$ By inspection to the form of $h$ in \eqref{phasespacevars}, consistency with $\Lambda$CDM also implies that $h(z=0)\approx1/2$, or equivalently, $H(z=0)\approx H_0$. 

\begin{figure}[htbp]
    \centering
    \includegraphics[width=\columnwidth]{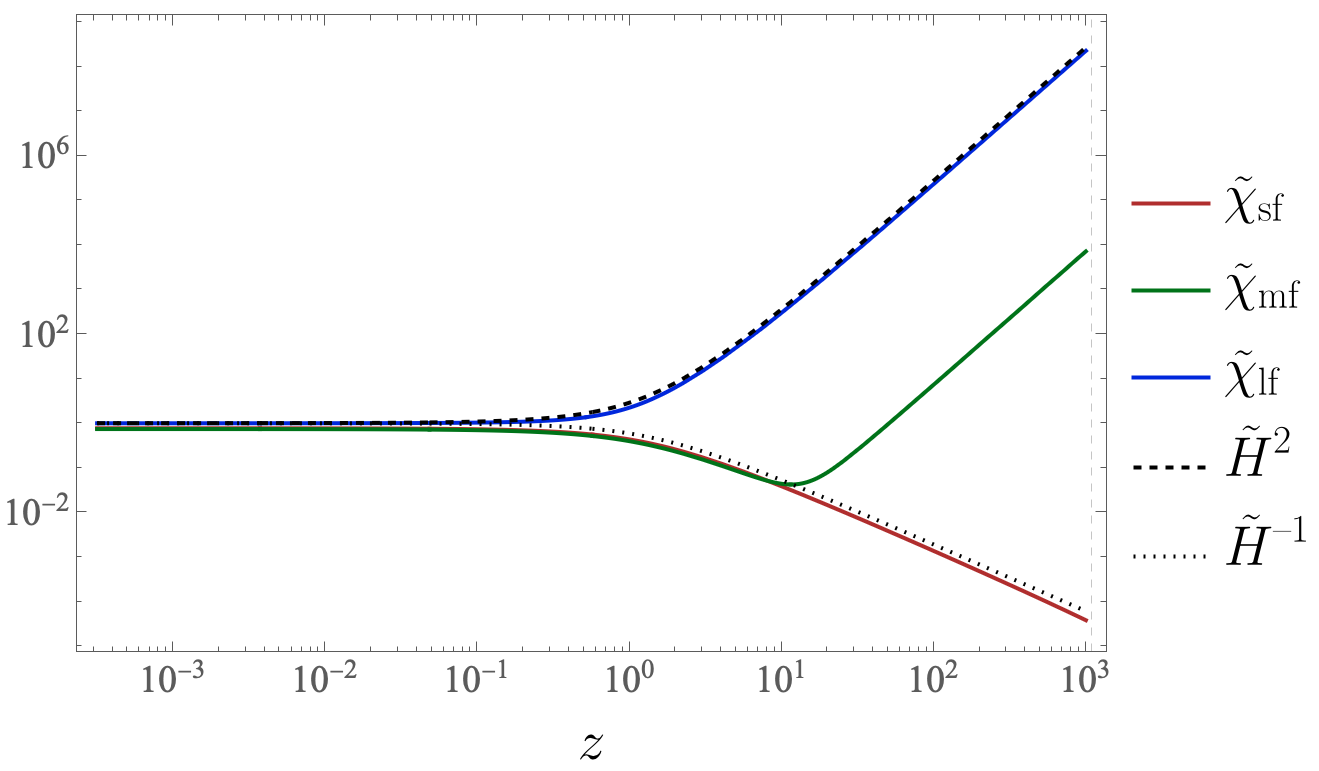}
    \caption{
    Representative solutions for small, middle, and large field initial conditions, compared to the universal profiles, where $\tilde{\x}=\x/\x_c$ and $\tilde{H}=H/H_0$. We observe the middle-field solution converging to the small-field solution around redshift $z_*\sim 10$. These solutions were integrated with the initial conditions and mass parameter described in Table \ref{table:ICs}. The phase space variables and equation of state evolution for these three specific field solutions are shown in Fig. \ref{fig:eos-branches}.
    }
    \label{fig:fieldtrajectories}
\end{figure}

\begin{figure*}[htbp]
    \centering
    \includegraphics[width=0.29\textwidth]{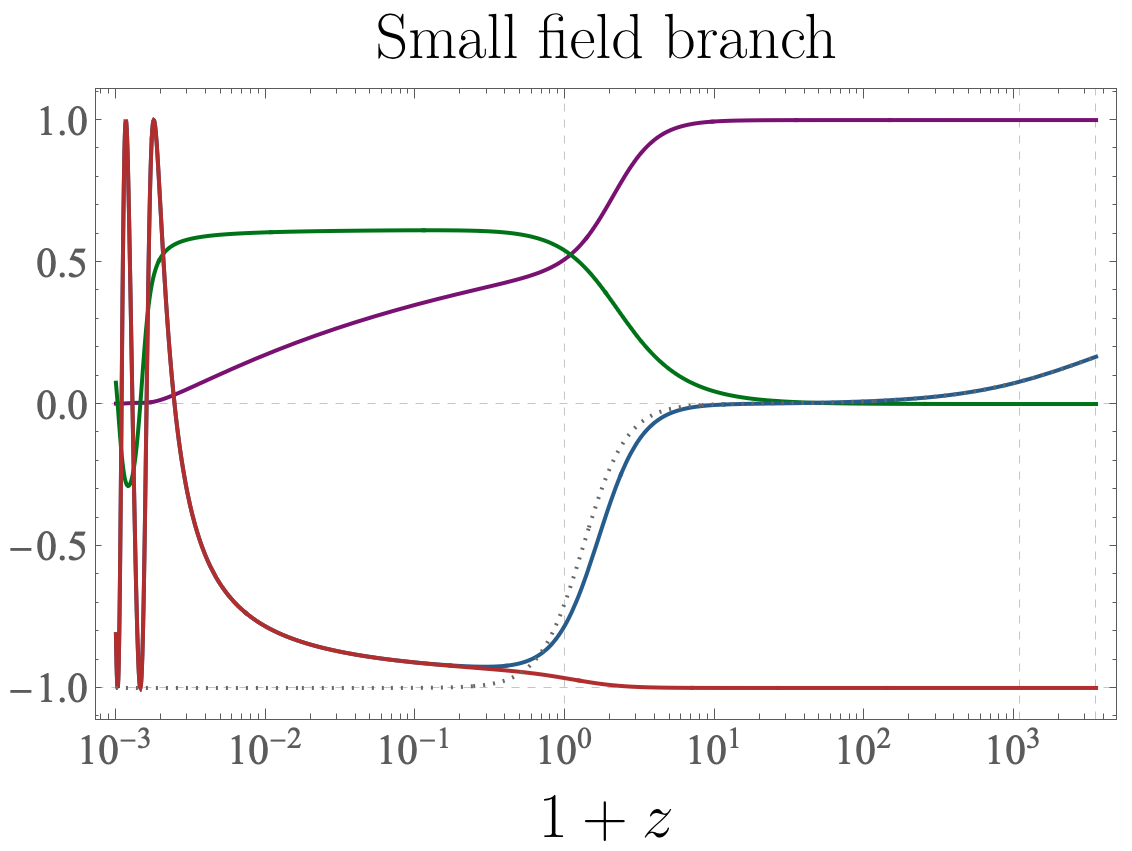}
    \hfill
    \includegraphics[width=0.29\textwidth]{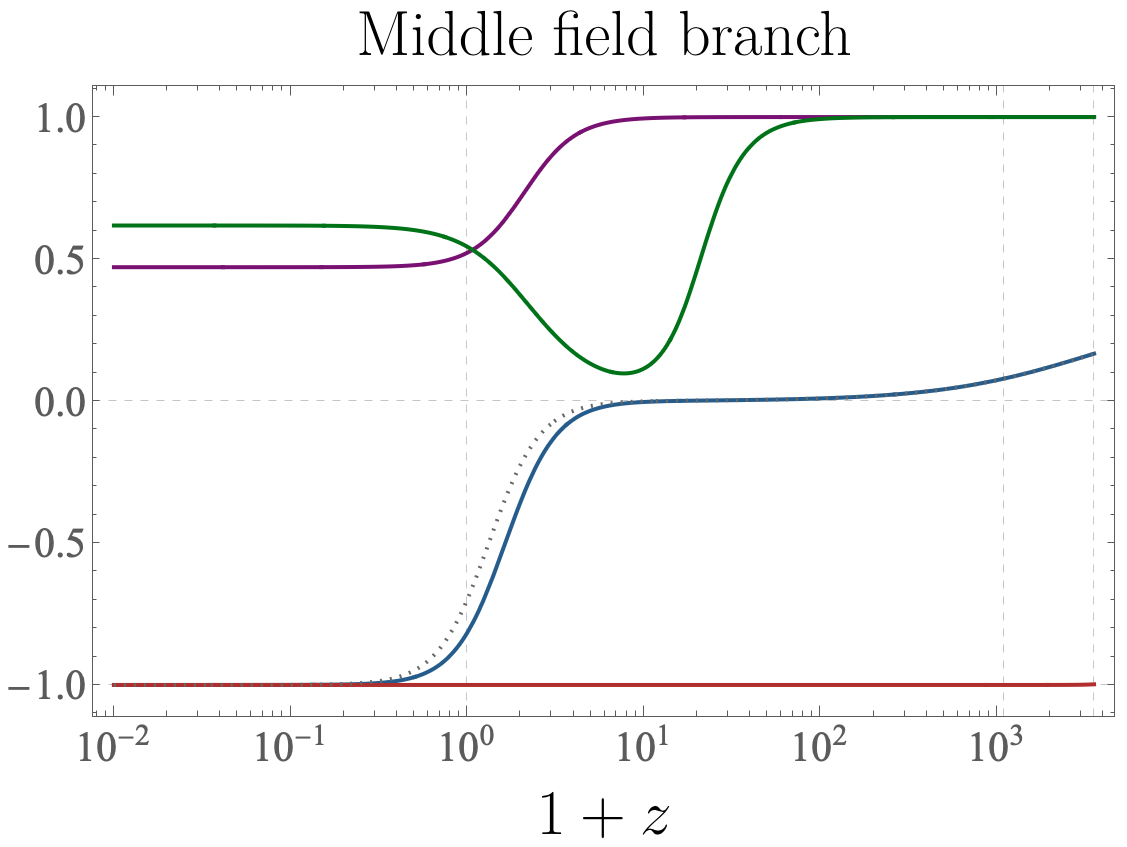}
    \hfill
    \includegraphics[width=0.36\textwidth]{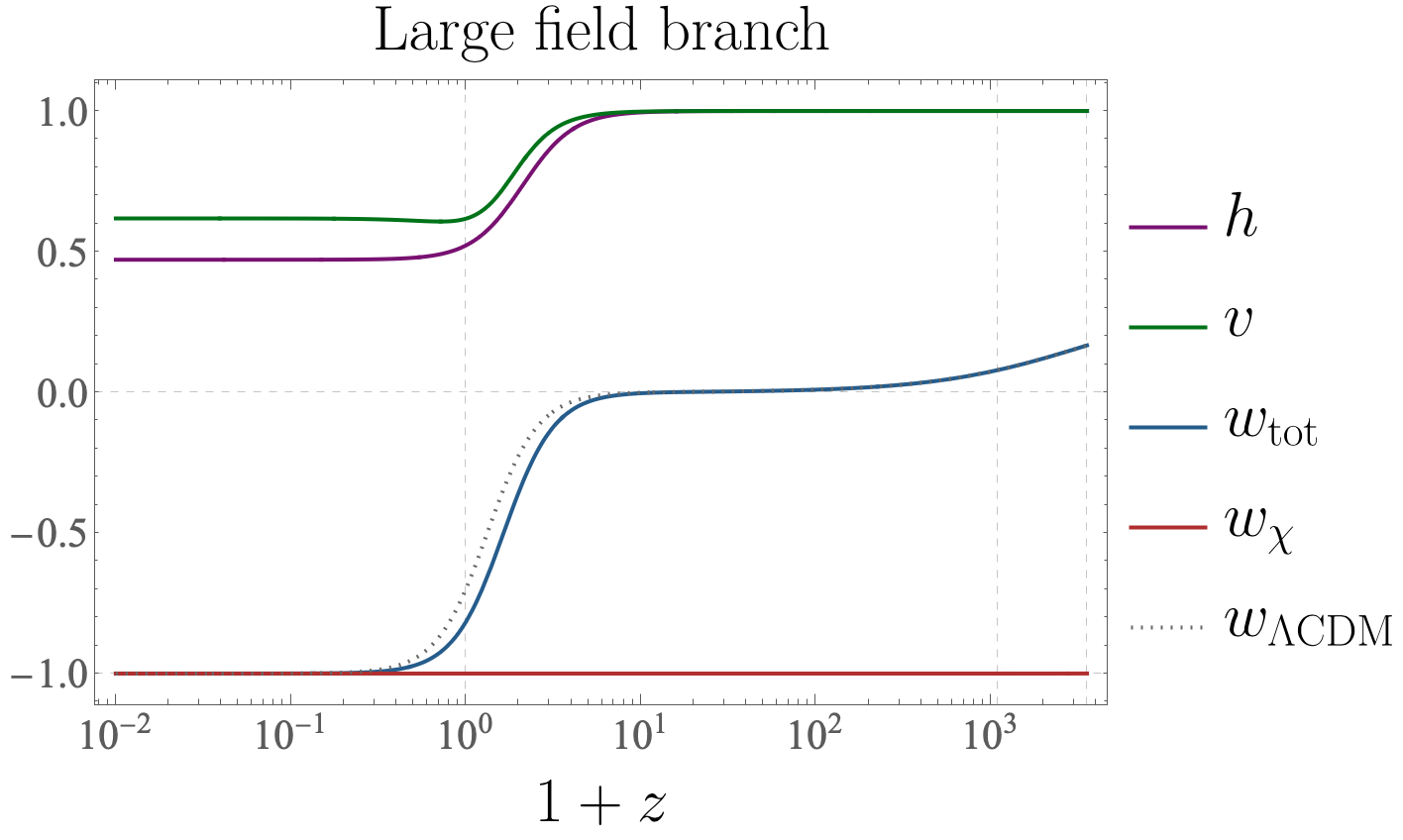}

    \caption{
    Representative cosmological trajectories for the three different branches of initial conditions shown in Table \ref{table:ICs}. We numerically observe two generic attractor behaviors in the limit as $z\rightarrow -1$. One attractor is de Sitter with $\chi\rightarrow \chi_c$, $w_{\infty}=-1$ and $H_{\infty}\approx 0.94H_0$. For the other, the 3-form EoS rapidly oscillates about $w=0$ such that $\langle w_{\infty}\rangle=0$, representing a matter-dominated attractor with $H_\infty\rightarrow0.$ The ultralight mass range that we choose, $m_\x/H_0\in[10^{-2},10^{-10}]$, results in the 3-form EoS behaving similarly to a cosmological constant up until the present. The phase space variable $h$ evaluated today for each trajectory is $h(z=0)\approx 0.5,$ corresponding to $H(z=0)\approx H_0$.
    }
    \label{fig:eos-branches}
\end{figure*}

Figure~\ref{fig:fieldtrajectories} shows the 
representative solutions for small, middle, and large field dynamics, 
with the parameters being described in Table.~\ref{table:ICs}. 
We confirm that the large- and small-field solutions are well 
described by Eqs.~\eqref{eq:lfanal} and \eqref{eq:sfanal}, respectively. The middle field solution in this example converges
on to the small field solution at $z \sim 10$. 
Figure~\ref{fig:eos-branches} shows the time evolutions of the corresponding phase space variables $v$ and $h$ as well as the equation of state, including the future dynamics $z\rightarrow -1$. 
The small-field solutions evolves into the matter-like oscillation, 
while the large- and middle-field solutions (in this example) approach the de Sitter attractor. 

\begin{table}[htbp]
\begin{tabular}{c|c|c|c }
 & $\log_{10} (M)$ &$\log_{10}(\x_i/\x_c)$  & $\Omega_{\x,\text{eq}}\times 10^{11}$  \\ \hline
 Small field & -0.8 &-20  &5.012    \\
 Middle field & -7 & 5.5  &5.017  \\
Large field  & -12 &10  &5.013   
\end{tabular}
\caption{Mass parameter value, initial field value at matter-radiation equality, and dark energy density at equality corresponding to the three field trajectories in Figs. \ref{fig:fieldtrajectories} and \ref{fig:eos-branches}. The choices of $M$ correspond to physical masses $m_\x/H_0\sim [10^{-2},10^{-9},10^{-13}]$. The initial condition for the Hubble parameter at equality for each trajectory is set to $H_{\text{eq}}=10^{10.25}H_0$. }
\label{table:ICs}
\end{table}

\section{Birefringent photon interactions}\label{sec:Interactions}

Having introduced 3-form cosmology and dark energy, we here move to the effective interactions with the photon, which is described by the usual Maxwell Lagrangian $\mathcal{L}_A=-\tfrac{1}{4}F_{\mu\nu}F^{\mu\nu}$. Before studying the dynamics stemming from specific interactions, let us discuss which candidate EFT-inspired operators may give birefringence. From a model building perspective, at the level of the photon equations of motion, we need operators that produce the cross-product $i\vec{k}\times\vec{\varepsilon}_\pm=\pm k\vec{\varepsilon}$ in order to get the $\pm$ in the dispersion relation. Thus, a contraction with the (three-dimensional) Levi-Civita symbol must appear. Moreover, the interaction must be quadratic in $A_\mu$. Those two conditions taken together effectively restrict us to interactions being linear in $C$ and quadratic in $A$ (up to dimension-6).

There are two dimension-4 interactions we may write down,
\begin{equation}
    C_{\mu\nu\rho}F^{\mu\nu}A^\rho, \ \ \ \ C_{\mu\nu\rho}\tilde{F}^{\mu\nu}A^\rho.
\end{equation}
For the second interaction, we can contract the Levi-Civita tensor on $\tilde{F}$ to build the vector dual $V^\mu=\epsilon^{\mu\nu\alpha\beta}C_{\nu\alpha\beta}$ to obtain
\begin{equation}
   V_\mu F^{\mu\nu}A_\nu.
\end{equation}
In the isotropic background, $V^\mu\propto(\x,0,0,0)$, reducing this interaction to $\sim\chi\dot{A}_iA^i$, which does not contribute opposite signs to the equations of motion and hence does not produce any birefringent effects. The $CFA$ operator, on the other hand, does produce birefringence as we will show below.
This operator breaks the electromagnetic gauge symmetry explicitly, but we will show later that it is phenomenologically acceptable.

Due to the antisymmetry of $C_{\mu\nu\rho}$ and $G_{\mu\nu\rho\sigma}$, there are no dimension-5 interactions that we can write down without introducing a third field. This is another distinguishing feature of cosmic birefringence from a 3-form compared to an axion; the axion-photon coupling is dimension-5.  

At dimension-6, we may write\footnote{One can also write down dimension-6 operators schematically of the form $C\nabla F F$ or $C\nabla F\tilde{F}$. By integrating by parts, anti-symmetrizing $\nabla C$ into $G$, and using the Bianchi identity for $F_{\mu\nu}$, it is straightforward to show such operators can be written in the form of \eqref{GFFops}.}
\begin{equation}\label{GFFops}
G_{\mu\nu\rho\sigma}F^{\mu\nu}F^{\rho\sigma}, \ \ \ G_{\mu\nu\rho\sigma}F^{\mu\nu}\tilde{F}^{\rho\sigma}.
\end{equation}
For the $GF\tilde{F}$ operator, using \eqref{Gcomps} for the components of $G_{0ijk}$, it is straightforward to show that the interaction reduces to $\sim(\dot{\x}+3H\x)\dot{A}_i\dot{A}^i$ at the level of the Lagrangian, which is effectively an artifact of two Levi-Civita symbols contracted together. This clearly will not produce a $\pm$ sign in the equations of motion, and therefore will not produce birefringence. The $GFF$ interaction indeed produces cosmic birefringence.

\subsection{Dimension-6 interaction}

We will first focus on the operator given explicitly by
\begin{equation}\label{Lint1}
    \mathcal{L}_{int}=-\frac{\epsilon}{4\Lambda^2}\F_{\mu\nu\rho\sigma}F^{\mu\nu}F^{\rho\sigma},
\end{equation}
where $\epsilon=\pm1$ and $\Lambda$ has dimensions of mass. The interaction respects the $U(1)_{EM}$ gauge invariance as well as the threefold gauge invariance of the 3-form. By varying the photon action with respect to $A^\mu$, we obtain
\begin{equation}
    \nabla_\nu F^{\mu\nu}=\frac{\epsilon}{\Lambda^2}\nabla_\nu\Big(\F^{\nu\mu\alpha\beta}F_{\alpha\beta}\Big).
\end{equation}

The field equation for the photon field \eqref{photonfield}, including our 3-form interaction \eqref{Lint1} reduces to 
\begin{equation}
    \ddot{A}_{k,\pm}+H\dot{A}_{k,\pm}+(\tfrac{k^2}{a^2}\pm \tfrac{k}{a}\alpha(t))A_{k,\pm}=0,
\end{equation}
where 
\begin{equation}
    \alpha(t)=\frac{\epsilon}{\Lambda^2}(\ddot{\chi}+3\frac{\mathrm{d}}{\mathrm{d}t}(H\chi)).
\end{equation}
The dispersion relation for the two helicity states in the WKB approximation is
\begin{equation}
    \omega_\pm\simeq \frac{k}{a}\pm \frac{\epsilon}{2\Lambda^2}(\ddot{\chi}+3\frac{\mathrm{d}}{\mathrm{d}t}(H\chi)), 
\end{equation}
such that the rotation angle is 
\begin{equation}\label{mainbeta}
    \beta=-\frac{\epsilon}{2\Lambda^2}\int_{t_{\text{LSS}}}^{t_0} \mathrm{d} t(\ddot{\chi}+3\frac{\mathrm{d}}{\mathrm{d}t}(H\chi)).
\end{equation}
Defining
\begin{equation}\label{X}
    X=\dot{\chi}+3H\chi,
\end{equation}
we see that the integral is simply
\begin{equation}\label{dim6generalbeta}
    \beta=-\frac{\epsilon}{2\Lambda^2}\int_{t_{\text{LSS}}}^{t_0}\mathrm{d}t \dot{X}=-\frac{\epsilon}{2\Lambda^2}\Big(X(t_0)-X(t_{\text{LSS}})\Big).
\end{equation}
Thus, in contrast to the axion case, the rotation angle is sensitive to the change in \textit{field velocity} as well as the field value and expansion history through the $H\chi$ contribution. 

Interestingly, using the equation of motion for $\x$ as in \eqref{chiEOM}, $\dot{X}=-dU/d\x,$ it follows that
\begin{equation}\label{beta4dUdx}
    \beta=\frac{\epsilon}{2\Lambda^2}\int_{t_{\text{LSS}}}^{t_0}dt\frac{dU}{d\x}.
\end{equation}
Thus, the birefringence angle from \eqref{Lint1} is highly sensitive to the shape of the potential, vanishing exactly when $U'=0,$ which is the case where $\x$ mimics a cosmological constant. We will discuss this point further in Sec. \ref{sec:ProfilesAndUniversality}.

By taking the number of $e$-folds, $N=-\ln(1+z)$, as the time variable, we have that $\dot{X}=H(\chi'+3\chi)$, where $\chi'=d\chi/dN$. Writing this in terms of the phase space variables $\chi=\chi_cf(v(N))$ and $H=H_0\tilde{f}(h(N))$ as in \eqref{phasespacevars} illustrates the scaling behavior for \eqref{dim6generalbeta}. That is, $$\Lambda^{-2}\dot{X}=\Lambda^{-2}H(\chi'+3\chi)=\Lambda^{-2}H_0\chi_c\tilde{f}(h)(f(v)+3f'(v)).$$ Recalling $\chi_c\propto M_p,$ we obtain $\beta\sim \frac{H_0M_p}{\Lambda^2}I_1(v,h)$.
Here, $I_1$ depends on the behavior of the compact phase space variables $v$ and $h$, which are in turn sensitive to the precise cosmological evolution. Thus, up to the factor of $I_1$,
\begin{equation}\label{dim6betascaling}
    \beta\propto\frac{H_0M_p}{\Lambda^2}\sim
    \frac{10^{-24}\text{ GeV}^2}{\Lambda^2}. 
\end{equation}
Note that while $f(v)$ and $\tilde{f}(h)$ are order of the unity today, they are large (or small for the small-field solution; see Sec.~\ref{subsec:analdyn3form}) at recombination, and hence $I_1$ does not have to be order unity.
For this interaction to produce $\beta\sim0.3^\circ,$ either $I_1$ must be enormous or $\Lambda$ must be very small, corresponding to an extremely large coupling. Explicitly, we have 
\begin{equation}\label{d6beta(z)}
    \beta(z)=-\frac{\epsilon H_0M_p}{\Lambda^2}\Big[I_1(v_z,h_z)-I_1(v_{\text{LSS}},h_{\text{LSS}})\Big],
\end{equation}
where $I_1(v,h)=\frac{1}{\sqrt{6}}\sqrt{\frac{h}{1-h}}[\frac{(1+v^2)v'+3v(1-v^2)}{(1-v^2)^2}]$ can be thought of as a function of redshift through $v_z=v(N(z))$ and $h_z=h(N(z))$, while $v_{\text{LSS}}$ and $h_{\text{LSS}}$ denote the values at the surface of last scattering. The value of $I_{1}(v,h)$ at a given redshift will slightly differ between the three branches discussed in Sec. \ref{sec:phasespace}.

\subsection{Dimension-4 interaction}\label{sec:dim4}

Here, we will turn our attention to the interaction
\begin{equation}\label{Lint2}
    \mathcal{L}_{int}=\epsilon\lambda C_{\mu\nu\rho}F^{\mu\nu}A^\rho,
\end{equation}
where $\lambda$ is the dimensionless coupling. This modifies the equation of motion for the photon as
\begin{equation}\label{FeomLint2}
    \nabla_\nu F^{\mu\nu}=\epsilon\lambda\Big[C_{\alpha\beta}^{ \ \ \mu}F^{\alpha\beta}+2\nabla_\alpha(C^{\mu\alpha\beta}A_\beta)\Big].
\end{equation}
The interaction breaks the $U(1)_{\text{EM}}$ gauge invariance, which we will discuss further momentarily. The dynamics for the Fourier modes are governed by 
\begin{equation}
    \ddot{A}_{k,\pm}+H\dot{A}_{k,\pm}+(\tfrac{k^2}{a^2}\pm \tfrac{k}{a}\epsilon\lambda\chi(t))A_{k,\pm}=0,
\end{equation}
such that
\begin{equation}\label{dim4dispersion}
    \omega_\pm\simeq\frac{k}{a}\pm \frac{\epsilon \lambda}{2}\chi
\end{equation}
in the WKB approximation. Thus, $\omega_+-\omega_-=\epsilon\lambda\chi(t)$, and we obtain
\begin{equation}\label{beta2a}
    \beta=-\frac{\epsilon\lambda}{2}\int_{t_{\text{LSS}}}^{t_0} dt\chi(t)=-\frac{\epsilon\lambda}{2}\int_{N_{\text{LSS}}}^0\frac{dN}{H(N)}\chi(N).
\end{equation}
The rotation angle here depends on the total integrated field evolution of the scalar $\chi$, where as in the axion case, $\beta$ only depends on the endpoint values of the ALP \eqref{axionbeta}.

Writing \eqref{beta2a} in terms of the phase space variables  \eqref{phasespacevars}, and recalling that $\chi_c=\sqrt{\frac{2}{3}}M_p$, we have 
\begin{equation}\label{d4beta(z)}
    \beta(z)=-\epsilon\lambda\Big(\frac{M_p}{H_0}\Big)\Big[I_2(v_z,h_z)-I_2(v_{\text{LSS}},h_{\text{LSS}})\Big],
\end{equation}
with $I_2=\tfrac{1}{\sqrt{6}}\int dN\frac{v}{1-v^2}\sqrt{\frac{1-h}{h}}$. 

For this case, we see a very different scaling compared to the dimension-6 interaction:
\begin{equation}\label{dim4betascaling}
    \beta\propto\lambda\frac{M_p}{H_0}\sim \lambda \times 10^{60}.
\end{equation}
Therefore producing $\beta\sim0.3^\circ,$ we either need the integral $I_2$ to be exceedingly small or we need the coupling $\lambda$ to be very small. 

The coupling reflects the hierarchy between the cosmological UV scale ($M_p$) and IR scale ($H_0$). Since the operator is mass dimension-4, we expect the scaling to appear either as $(M_p/H_0)^{\pm 1}$. For the coupling $\lambda$ to be very small, consistent with effective field theory expectations, the scaling should appear as in \eqref{dim4betascaling}. Moreover, as we will argue below, there is an elegant cosmological consequence of this operator being responsible for cosmic birefringence, in that it demands the photon has a mass within a few orders of magnitude of the Hubble scale.

\vspace{5mm}

\textit{Restoring gauge invariance.}
The dimension-4 interaction \eqref{Lint2} explicitly breaks $U(1)_{\rm EM}$ gauge invariance through the presence of the photon field $A_\mu$: the transformation 
\begin{equation}\label{gauge1}
    A_\mu\to A_\mu+\nabla_\mu\alpha 
\end{equation}
explicitly induces the vertex 
\begin{equation}\label{alphavertex}
    \mathcal{L}_g\supset -\lambda C_{\mu\nu\rho}F^{\mu\nu}\nabla^\rho\alpha .
\end{equation}
It is therefore possible that electric charge is no longer conserved.

Let us first check the conservation of the current associated to the RHS of \eqref{FeomLint2},
\begin{equation}
J^\mu_{(C)}\propto C_{\alpha\beta}^{ \ \ \mu}F^{\alpha\beta}+2\nabla_\alpha(C^{\mu\alpha\beta}A_\beta),
\end{equation}
and inspect $\nabla_\mu J^\mu_{(C)}.$ The second term vanishes by antisymmetry, such that 
\begin{equation}
\nabla_\mu J^\mu_{(C)}
\propto\nabla_\mu\left(C^{\mu\alpha\beta}F_{\alpha\beta}\right).
\end{equation}
It happens to be the case that this vanishes on the background. Using \eqref{Ccomps}, we see that 
\begin{equation}
    \nabla_i \left(a^{-3} \chi\, \epsilon^{ijk} F_{jk} \right) = 0.
\end{equation}
In terms of the magnetic field, $B^i = \frac{1}{2}\epsilon^{ijk}F_{jk}$, this becomes
\begin{equation}\label{constraint2}
    \nabla \cdot (\chi\, \mathbf{B}) = 0.
\end{equation}
The 3-form scalar is spatially isotropic on the background, thus $\nabla\chi=0$, and  $\nabla\cdot\mathbf{B}=0$ follows from the Bianchi identity. However, as soon as we introduce perturbations of the form $\chi=\chi_0(t)+\delta\chi(\mathbf{x},t)$, \eqref{constraint2} will no longer be satisfied. 

Consequently, conservation of charge necessitates the introduction of a longitudinal degree of freedom. To see this, we introduce the gauge invariant quantity
\begin{equation}
    A_\mu \rightarrow A_\mu + \tfrac{1}{m_\gamma}\nabla_\mu \pi,
\end{equation}
where $\pi$ is a mass dimension-1 scalar St\"uckelberg field\footnote{See \textit{e.g.} \cite{Ruegg:2003ps} for a review of the St\"uckelberg mechanism.} transforming as
\begin{equation}\label{gauge2}
    \pi \to \pi - m_\gamma\alpha.
\end{equation}
Here $m_\gamma$ is the induced photon mass. Under this replacement, the interaction \eqref{Lint2} becomes
\begin{equation}
    \mathcal{L}_{\rm int}
    =-\lambda 
    C_{\mu\nu\rho} F^{\mu\nu} \left(A^\rho + \tfrac{1}{m_\gamma}\nabla^\rho \pi \right),
\end{equation}
with $\mathcal{L}_g$ from \eqref{alphavertex} vanishing. 
The first term reproduces the original interaction, while the second results in a coupling between the 3-form, photon field strength, and the gradient of the St\"uckelberg field,
\begin{equation}\label{StuckelbergInt}
    \mathcal{L}_{\pi}
    =-\frac{\lambda}{m_\gamma}
    C_{\mu\nu\rho} F^{\mu\nu} \nabla^\rho \pi.
\end{equation}
Introducing a kinetic term and preserving gauge invariance necessitates the introduction of the mass term for the photon. The Lagrangian
\begin{equation}
\begin{split} 
\mathcal{L}_{GI}
&=
-\frac{1}{4} F_{\mu\nu} F^{\mu\nu}
-\frac{1}{2} m_\gamma^2 (A_\mu + \tfrac{1}{m_\gamma}\nabla_\mu \pi)(A^\mu + \tfrac{1}{m_\gamma}\nabla^\mu \pi) \\
&-\lambda C_{\mu\nu\rho} F^{\mu\nu} (A^\rho + \tfrac{1}{m_\gamma}\nabla^\rho \pi)
 - A_\mu J^\mu_\mathrm{matter}
\end{split}
\end{equation}
is manifestly invariant under the transformations \eqref{gauge1} and \eqref{gauge2}. Here, we have also explicitly introduced an electromagnetic  
current carried by the matter field, $J^\mu_\mathrm{matter}$.

Varying w.r.t. $A_\mu$ gives the photon equation of motion
\begin{equation}\label{conservedJ}
\begin{split}
J^\mu_\mathrm{matter}&= 
\nabla_\nu F^{\nu\mu}
- m_\gamma^2 (A^\mu+m_\gamma^{-1}\nabla^\mu\pi) \\
&
+2\lambda \nabla_\nu\left(C^{\nu\mu\rho} (A_\rho+m_\gamma^{-1}\nabla_\rho\pi)\right)
-\lambda C^{\mu\nu\rho} F_{\nu\rho},
\end{split}
\end{equation}
while the variation w.r.t. $\pi$ yields  
\begin{equation}
m_\gamma^2\nabla_\mu(A^\mu+m_\gamma^{-1}\nabla^\mu\pi)
=-\lambda
\nabla_\rho\left(C^{\mu\nu\rho} F_{\mu\nu}\right).
\label{eq:pieom}
\end{equation}
Taking the divergence of \eqref{conservedJ}, using antisymmetry and simplifying leaves us with the conservation of the current
\begin{equation}
\begin{split} 
\nabla_\mu J^\mu_\mathrm{matter} &= - m_\gamma^2 \nabla_\mu(A^\mu+m_\gamma^{-1}\nabla^\mu\pi) - \lambda \nabla_\mu\!\left(C^{\mu\nu\rho} F_{\nu\rho}\right) \\ &=0, 
\end{split}
\end{equation}
where we have used
the equation of motion for $\pi$.

The introduction of $\pi$ allows the additional, longitudinal degree of freedom to be incorporated while maintaining manifest gauge invariance. The role of $\pi$ may also be understood directly from the structure of the mixing term. Expanding the mass term gives a contribution proportional to $A^\mu \nabla_\mu \pi$, which couples the vector field to derivatives of the scalar. In momentum space, this interaction is proportional to $k_\mu A^\mu$, and therefore involves only the longitudinal component of the vector field. Transverse modes, satisfying $k_\mu \epsilon^\mu = 0$, do not couple to $\pi$ through this term.

This separation of longitudinal and transverse sectors has an important consequence for photon propagation. Birefringence arises from helicity-dependent modifications of the dispersion relation for the transverse photon polarizations. Since the St\"uckelberg field couples only to the longitudinal component, it does not directly affect the transverse modes responsible for birefringence. In particular, the derivative mixing term vanishes identically when evaluated on transverse polarization states.

As a result, the St\"uckelberg completion restores gauge invariance and ensures charge conservation without altering the transverse dispersion relations at leading order. The birefringence effects discussed in Eqs. \eqref{dim4dispersion}-\eqref{dim4betascaling} therefore continue to arise entirely from the original $C_{\mu\nu\rho}F^{\mu\nu}A^\rho$ interaction, while the St\"uckelberg sector modifies only the longitudinal dynamics of the vector field.

We here wish to comment about the relationship between the dimensionless coupling $\lambda$, the photon mass $m_\gamma$ and $H_0$. We see from \eqref{StuckelbergInt} that we do not have a smooth limit in taking the photon mass to zero. On the other hand, from an EFT perspective, the only UV scale in the problem is $M_p$. We thus argue that it is natural to write $\lambda$ as 
\begin{equation}\label{lambdamoverMp}
    \lambda\sim\frac{m_\gamma}{M_p},
\end{equation}
such that
\begin{equation}
    \lim_{m_\gamma\rightarrow 0}\lambda 
    C_{\mu\nu\rho} F^{\mu\nu} \left(A^\rho + \tfrac{1}{m_\gamma}\nabla^\rho \pi \right)=\frac{1}{M_p}C_{\mu\nu\rho}F^{\mu\nu}\nabla^\rho\pi,
\end{equation}
which a finite, Planck-suppressed, gauge-invariant quantity. The full theory in the decoupling limit is thus
\begin{equation}
    \begin{split}
        \mathcal{L}_{\text{dec}}&= -\frac{1}{4}F_{\mu\nu}F^{\mu\nu}-\frac{1}{48}G_{\mu\nu\rho\sigma}G^{\mu\nu\rho\sigma}-U(C^2) \\
        &-\frac{1}{2}(\nabla_\mu\pi)^2-\frac{1}{M_p}C_{\mu\nu\rho}F^{\mu\nu}\nabla^\rho\pi.
    \end{split}
\end{equation}
In order to see how the presence of the extra scalar d.o.f. affects the background dynamics, we inspect the following equations of motion conveniently in terms of the scalar $\chi$ and magnetic field $\mathbf{B}$: 
\begin{equation}\label{inhomo1}
    \nabla_\mu F^{\mu i}=-\frac{2a^3}{M_p}\nabla_j(\chi\epsilon^{jik}\nabla_k\pi),
\end{equation}
\begin{equation}\label{inhomo2}
    \Box\pi=-\frac{a^3}{M_p}\vec{\nabla}(\chi\mathbf{B}),
\end{equation}
\begin{equation}\label{inhomo3}
    \ddot{\chi}+3\frac{d}{dt}(H\chi)+U^\prime(\chi) =-\frac{a^3}{M_p}\mathbf{B}\cdot\vec{\nabla}\pi.
\end{equation}
This illustrates that the field equations are only modified in the presence of spatial inhomogeneities in the scalar fields. 

One can imagine that \eqref{inhomo1}-\eqref{inhomo3} may have an observational effect on e.g. galactic magnetic fields. If we expand the scalars as $\x=\bar{\x}(t)+\delta\x(x,t), \ \pi=\bar{\pi}(t)+\delta\pi(x,t)$, then the lowest order Maxwell source from \eqref{inhomo1} is 
$$J_{\text{inhom}}^i=-\frac{2a^3}{M_p}\epsilon^{ijk}(\nabla_j\delta\chi)(\nabla_k\delta\pi), $$
which is Planck suppressed, gradient suppressed, and second order in perturbations. We expect such a small source would not produce any observable effects. The homogeneous modes $\bar{\pi}$ and $\bar{\x}$ have respective sources
$$-\frac{a^3}{M_p}\langle\mathbf{B}\cdot\vec{\nabla}\delta\x\rangle, \ \ \ -\frac{a^3}{M_p}\langle\mathbf{B}\cdot\vec{\nabla}\delta\pi\rangle,$$ 
where the brackets denote spatial averaging. Therefore in principle, astrophysical regions where there is a strong magnetic field aligned with inhomogeneous gradients of the perturbations which survive statistical averages, the background scalars could have local, Planck suppressed sources. Such small effects will not significantly change the background evolution, and have no appreciable affect on photon birefringence.

In addition to a smooth decoupling limit, writing $\lambda$ as \eqref{lambdamoverMp} results in the following cosmological connection. In order to describe the observed cosmic birefringence, we saw generically the scaling 
\begin{equation}\label{photon&H0}
\beta\propto\lambda\frac{M_p}{H_0}\sim \frac{m_\gamma}{H_0},
\end{equation} 
up to the factor of the integral $I_2$ stemming from the 3-form dark energy dynamics. 

\subsection{Numerical results for massive potential}\label{sec:NumericalResultsForMassivePotential}

In section \ref{sec:3formDE}, we obtain three cosmological histories consistent with $\Lambda$CDM up to low redshift for the case of a massive (ultralight) 3-form, with $U(\x)=\tfrac{1}{2}m_\x \x^2.$ This choice of potential is such that $\beta_{\text{dim-6}}\propto \beta_{\text{dim-4}}$. In equation \eqref{beta2a}, we saw that $\beta_{\text{dim-4}}\propto\int dt\x$, while, using the equations of motion for the free 3-form, we found $\beta_{\text{dim 6}}\propto\int dtU'\sim\int dt\x$. For any other choice of potential, $\beta_{\text{dim-4}}$ and $\beta_{\text{dim-6}}$ will be different forms. 

Because of this, the expressions we call $I_1$ and $I_2$ are equal. Their values for the three different solution branches are 
\begin{equation}
    \begin{split}
        \text{Small field: } \ \ \  & \ \ \ 1.874\times 10^{-1}, \\
        \text{Middle field: }\ \ \ & \ \ \ 1.396, \\
        \text{Large field: }\ \ \ & \ \ \ 4.626\times 10^3.
    \end{split}
\end{equation}
Accounting for the $I_1=I_2\equiv I$, we have 
\begin{equation}
    \beta_{\text{dim-6}}=\frac{H_0M_p}{\Lambda^2}I, \ \ \ \ \ \beta_{\text{dim-4}}=\lambda\frac{M_p}{H_0}I\sim\frac{m_\gamma}{H_0}I,
\end{equation}
following our arguments for the introduction of the photon mass in Sec. \ref{sec:dim4}. Using $H_0M_p\sim 10^{-24}\text{ GeV}^2,$ the coupling for the dimension-6 operator must be
\begin{equation}
  10^{-24}\text{ GeV}^2 \lesssim\Lambda^2\lesssim 10^{-20}\text{ GeV}^2 
\end{equation}
in order for the interaction to explain cosmic birefringence. Requiring such an enormous coupling $\Lambda^{-2}$ is not consistent with the standard philosophy of effective field theory. We thus argue \eqref{Lint1} can not explain cosmic birefringence. 

For the dimension-4 interaction, we instead find that
\begin{equation}
    10^{-1}\lesssim\frac{m_\gamma}{H_0}\lesssim 10^3
\end{equation}
in order for \eqref{Lint2} to explain cosmic birefringence. Once the interaction is completed in a gauge-invariant manner through the St\"uckelberg mechanism, demanding both a smooth decoupling limit and the observed birefringence angle ties the photon mass directly to the gravitational infrared scale. This value lies many orders of magnitude below existing experimental sensitivity \cite{ParticleDataGroup:2024cfk}.

\section{Birefringence angle profiles and universality}\label{sec:ProfilesAndUniversality} 

In the previous section, we considered specifically whether or not our interactions can explain the cosmic birefringence of the CMB, and obtained approximations of the couplings required to achieve $\beta=0.3^\circ$. 
Here we will look at the redshift evolution of $\beta$, 
which may be detectable in the observation of radio galaxies at low redshifts, 
motivated by the study in Ref.~\cite{Naokawa:2025shr}. 
Importantly, Ref.~\cite{Naokawa:2025shr} showed that the redshift profile of cosmic birefringence induced by an ALP is universal and insensitive to the shape of the potential, 
as long as the ALP slow rolls from  recombination. 
Based on the discussion in Sec.~\ref{subsec:analdyn3form}, 
we will find the universal profiles for the dimension-4 operator 
in the case of the small and large-field solutions, 
while the profile strongly depends on the potential for the dimension-6 operator. 
We also compare them with the numerically evaluated profile of $\beta$ 
for the specific case of $U(\chi)=\tfrac{1}{2}m_\x\x^2$. 

As discussed in Sec.~\ref{subsec:analdyn3form}, 
we have two representative 3-form field evolutions as
\begin{align}
    \chi(t) &\simeq \left\{ \begin{array}{ll} \chi_c \exp[3(N(t)-N(t_0))/2] & \text{small field solution} \\ \\
    \chi_c \exp[-3(N(t)-N(t_0))]  & \text{large field solution}
    \end{array}\right.  \notag \\
    &\simeq \left\{ \begin{array}{ll} \chi_c (H(t)/H_0)^{-1} & \text{small field solution} \\ \\
    \chi_c (H(t)/H_0)^2  & \text{large field solution}
    \end{array}\right. , 
\end{align}
where the change of the dynamics when the 3-form field energy becomes non-negligible is omitted. 

The cosmic birefringence from the dimension-4 operator integrated up to a redshift $z$ 
is now evaluated as 
\begin{align}
    \beta(z) &= -\frac{\epsilon \lambda}{2} \int_{t(z)}^{t_0} \mathrm{d} t \chi(t) \notag \\
    &= -\frac{\epsilon \lambda}{2} \int_0^z \mathrm{d}z'  \frac{\chi(z')}{(1+z')H(z')} \notag \\
    &= -\frac{\epsilon \lambda \chi_c}{2} \times \left\{\begin{array}{ll}
    \int_0^z  \frac{H_0}{(1+z')H(z')^2} & \text{small field solution} \\ \\
     \int_0^z  \frac{H(z')}{(1+z')H_0^2} & \text{large field solution}
    \end{array}
    \right. 
\end{align}
Approximating that the Hubble parameter is the same as for  $\Lambda$CDM,
\begin{equation}
    H(z)= H_0\sqrt{(1-\Omega_{\Lambda 0}) (1+z)^3 + \Omega_{\Lambda 0}}, 
\end{equation} 
the profile reads 
\begin{align}
    \beta(z) & = -\frac{\epsilon \lambda}{2} \frac{\chi_c}{H_0} F_1(z) \notag \\
    &= \beta(z_\mathrm{LSS}) \frac{F_1(z)}{F_1(z_\mathrm{LSS})}, \notag 
\end{align}
where we have defined
\begin{align}
    F_1(z) & \equiv \int_0^z \mathrm{d} z' \frac{1}{(1+z')^4 (1-\Omega_{\Lambda 0})  + (1+z')\Omega_{\Lambda 0}} \label{Universal_dim4_smallfield}
\end{align}
for the small-field solution, 
where $\beta(z)$ starts to move around $z \sim \mathcal{O}(1)$ 
and can be the target for the test with radio galaxies. 
We find that it is potential independent and, remarkably, it is exactly the same to the ALP universal profile~\eqref{ALPUnivProfile}. 

For the large-field solution, we obtain
\begin{align}
    \beta(z) & = -\frac{\epsilon \lambda}{2} \frac{\chi_c}{H_0} F_2(z) \notag \\
    &= \beta(z_\mathrm{LSS}) \frac{F_2(z)}{F_2(z_\mathrm{LSS})}, \notag 
\end{align}
where we have defined
\begin{align}
    F_2(z) & \equiv \int_0^z \mathrm{d} z' \frac{\sqrt{(1-\Omega_{\Lambda 0}) (1+z')^3 + \Omega_{\Lambda 0}}}{(1+z')},  \label{Universal_dim4_largefield}
\end{align}
which is, once more, potential independent. 
The profile damps at $z\simeq z_\mathrm{LSS}$
and hence we would not see cosmic birefringence from radio galaxies, 
as is the case of ALP dark matter. 
This case 
would be tested by the CMB EB power spectrum 
with the absence of the signal at low-$\ell$ as well as 
the distortion at $\ell \sim 10^3$
in the forthcoming CMB observations~\cite{Nakatsuka:2022epj,Carralot:2026kps}, 
which requires more detailed examination. 

Since the middle-field solution is a superposition of the small- and large-field, the birefringence angle is also a superposition,
\begin{equation}
    \beta_{\text{mf}}\simeq\mA\beta_{\text{sf}}+\mB\beta_{\text{lf}}.
\end{equation}
The shape of the normalized solution for $\beta_{\text{mf}}$ is therefore sensitive to the ratio $\mA/\mB$. From \eqref{ABratio}, we also know that the ratio $\mA/\mB$ is related to the redshift at which point $\x_{\text{mf}}$ converges to the small-field attractor. Since the values of $\mA$ and $\mB$ simply correspond to the initial conditions for the field, we conclude that although $\beta(z)$ is indeed independent of the choice of 3-form potential, it is sensitive to the initial field configuration. 

On the other hand, the dimension-6 operator gives a potential dependent profile. 
The cosmic birefringence integrated up to a redshift $z$ is given by
\begin{align}
    \beta(z) &= -\frac{\epsilon}{2 \Lambda^2} \int_{t(z)}^{t_0} \mathrm{d} t ({\dot \chi+ 3 H \chi})^\cdot = \frac{\epsilon }{2 \Lambda^2} \int_{t(z)}^{t_0} \mathrm{d} t U' \notag \\
    & = -\frac{\epsilon }{2 \Lambda^2} \int_0^z\mathrm{d}z' \frac{ U'}{(1+z') H(z')},  
\end{align}
which is $U'$ dependent. This is to be expected, following the discussion around \eqref{beta4dUdx}.
In the case of ALP dark energy, the scalar field slow-rolls even before the dark energy
domination and $U'$ can be approximated to be constant, 
which is the reason why the $\beta$ profile is potential independent. 
In the 3-form case, $\chi$ field does not slow roll, and $U'$= const. approximation
does not work. 
Therefore we conclude that the cosmic birefringence from the dimension-6 operator 
is in general potential dependent. 

\begin{figure}[t]
  \centering

  \includegraphics[width=\columnwidth]{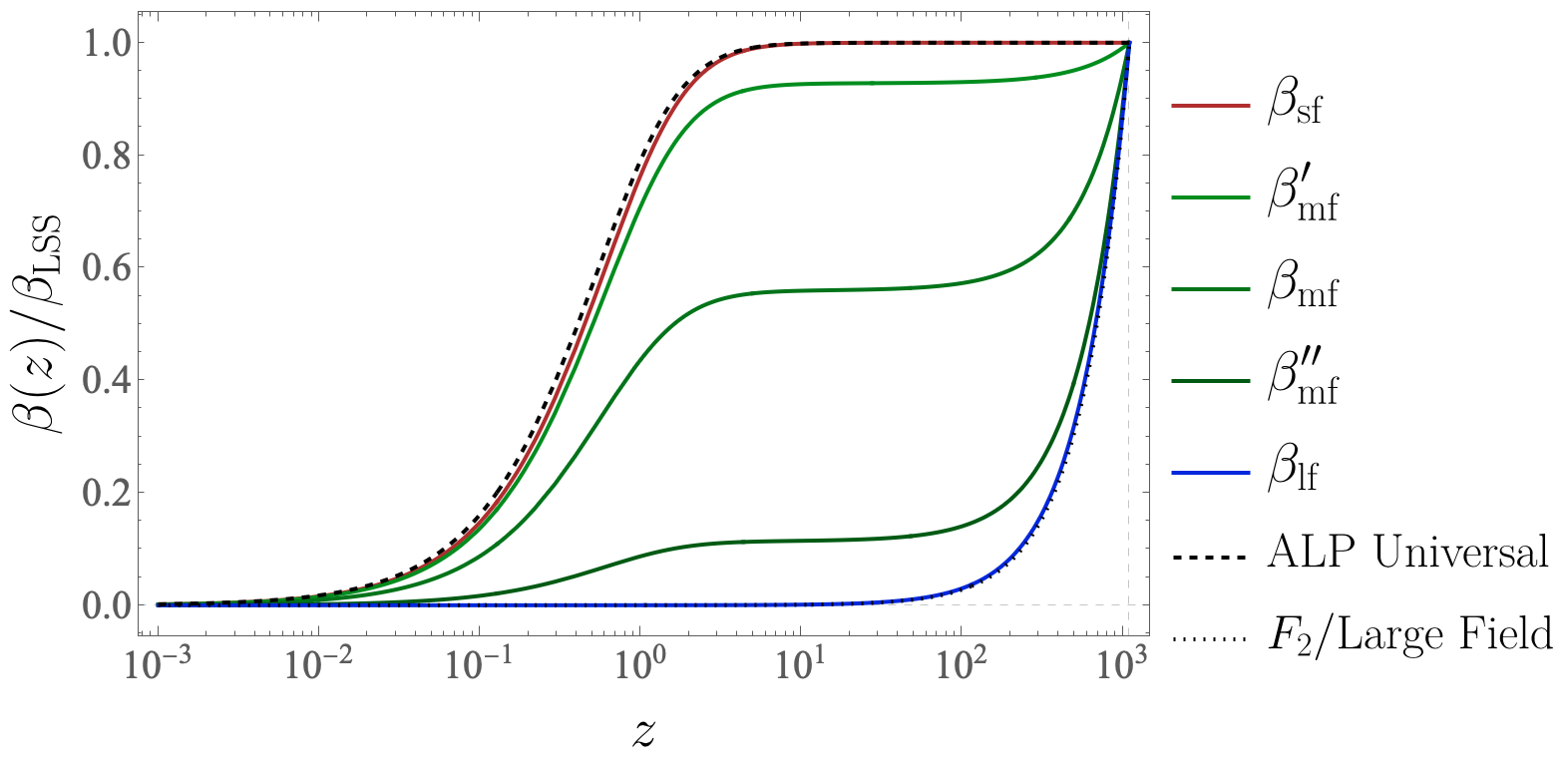}

  \vspace{0.5em}

  \includegraphics[width=\columnwidth]{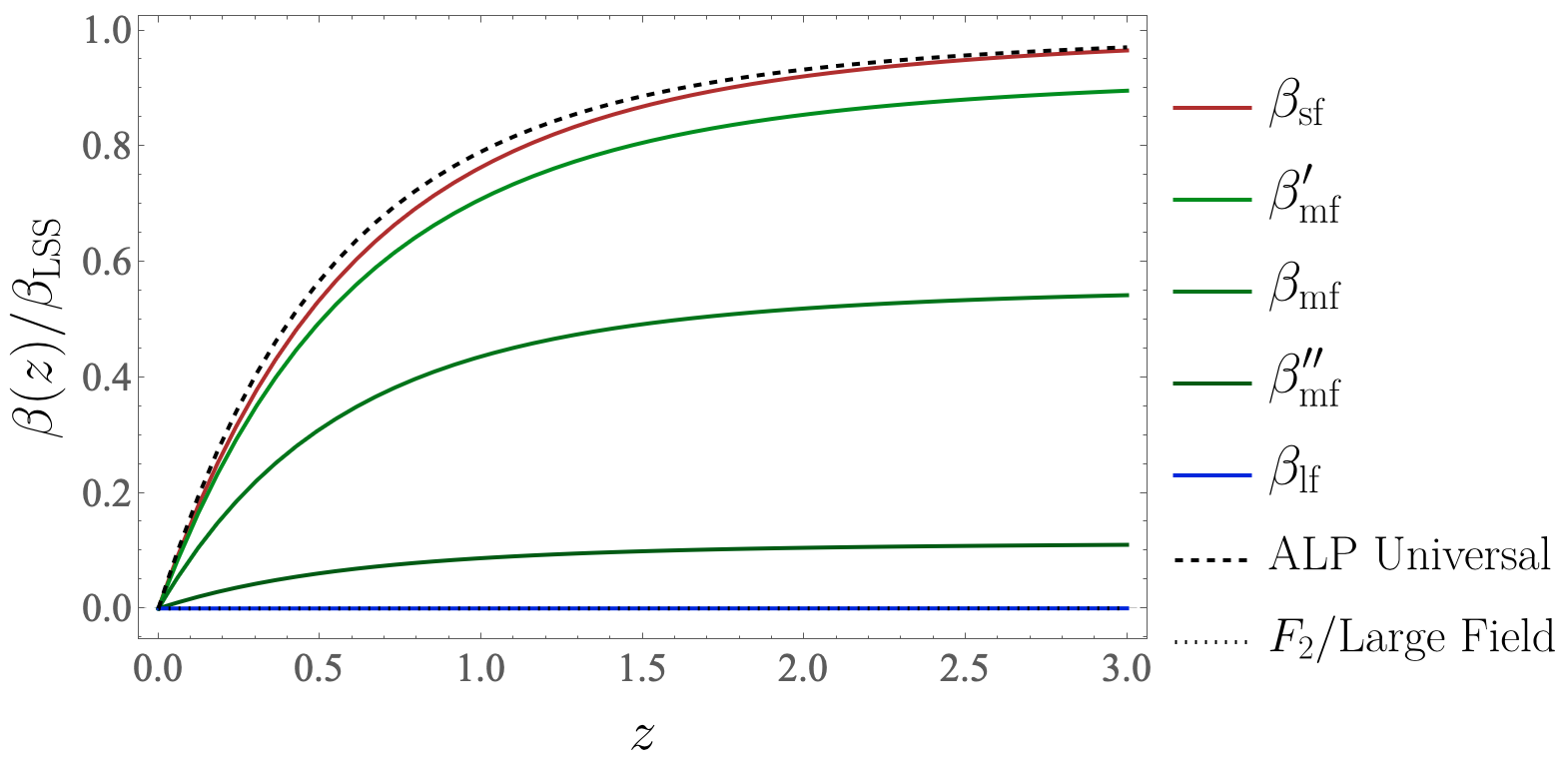}

  \caption{Comparison of profiles for the rotation angles for our universal results, \eqref{Universal_dim4_smallfield} and \eqref{Universal_dim4_largefield}, to the numerically integrated results for the field evolutions for ultralight, massive 3-form fields, for $\beta(z)\sim\int dt\x(t)$. All profiles are normalized to $\beta_{\text{LSS}}=\beta(z_{\text{LSS}})$. The small, middle, and large field initial conditions are the same as those from Table \ref{table:ICs}. We have additionally plotted two middle field solutions with initial conditions $\log(\x_i/\x_c)=4.5,6.5$, and mass parameters $\log(M)=-6,-8,$ respectively, for the curves $\beta'_{\text{mf}}$ and $\beta''_{\text{mf}}$. }
  \label{fig:Profiles}
\end{figure}

Figure~\ref{fig:Profiles} shows results comparing the universal profiles to those obtained from integrating $\sim\int dt\chi(t)$ for the numerical solutions corresponding to the small and large field trajectories, including three examples of middle field trajectories. As we discussed in Sec. \ref{sec:NumericalResultsForMassivePotential}, choosing a massive potential results in $\beta_{\text{dim-6}}\propto\beta_{\text{dim-4}}$. We see that the family of middle-field initial conditions essentially extrapolates between the universal profiles for the small- and large-field cases.

\section{Discussion and Conclusion}\label{sec:Conclusion}
Since an ALP is arguably the simplest explanation of photon birefringence, confirmation of isotropic cosmic birefringence in the CMB will be highly suggestive of the existence of an ALP. However, it is also a possibility that the plethora of direct and indirect searches for ALPs \cite{Marsh:2015xka} continue to return null results, necessitating alternative explanations. We have shown here that 3-form dark energy can either mimic ALP cosmic birefringence or show significant departures from the ALP case.

Up to dimension-6, only two unique operators produce isotropic cosmic birefringence. The dimension-6, gauge invariant operator $\sim GFF$ necessitated a coupling of $O(10^{23})$ GeV$^{-2}$ to explain $\beta\sim 0.3^\circ$. From an EFT perspective, the coupling $\Lambda^{-2}$ should be small, therefore we argue the $GFF$ operator cannot be responsible for the CMB cosmic birefringence. The dimension-4 operator $\sim CFA$, on the other hand, is a possible candidate, at the expense of exciting the longitudinal mode of the photon. We further argue that explaining $\beta\sim 0.3^\circ$ and demanding a smooth decoupling limit requires the photon mass to be within a few orders of magnitude of the Hubble scale $m_\gamma\sim H_0\sim 10^{-33}$ eV, c.f. \eqref{photon&H0}. Not only is this within current experimental constraints, it is interesting from a fundamental physics point of view in that the photon mass is directly tied to the gravitational IR scale. 

From a purely theoretical perspective, it is natural to ask if an effective theory involving $p$-form fields has a string theory origin. For the two interactions we studied in this work, the short answer is that there is no obvious stringy origin. Regarding the dimension-4 interaction $\sim CFA$, it is possible that this may stem from the Wess-Zumino term $\mathcal{L}\sim C\wedge e^{F-B}$, where $B$ is a 2-form and $C$ is a sum over $p$-forms from $p=0,...,9$ \cite{Green:1996bh}. Expanding the exponential and integrating out the $B$-field may produce a term of the form $CFA$, but its emergence certainly is not clear. On the other hand, Chern-Simons-like forms $\sim cF\wedge A$ (where $c$ is a constant) appear as topological response terms in the fractional quantum Hall effect \cite{PhysRevLett.62.82}. Our $CFA$ interaction potentially may be understood from a similar viewpoint. We stay agnostic to the origins of the interactions studied here and treat them simply from an EFT perspective.

We restricted our attention to a massive potential for the 3-form field. Such a setup can be consistent with $\Lambda$CDM, and allow for deviations in the EoS at low redshift. The dynamical dark energy scenario is currently mildly preferred over $\Lambda$CDM following recent analysis by DESI and eBOSS \cite{DES:2026jmi,DESI:2024lzq,DESI:2025zgx,eBOSS:2020yzd}, although the community is still far from a definitive conclusion on this exciting possibility.  



Our results here are novel in the sense that our approximations of the couplings required to explain the CMB birefringence are fully consistent with viable cosmological histories, however, a simple, massive potential is not the only choice that can be broadly consistent with $\Lambda$CDM as dynamical dark energy \cite{Koivisto:2009ew}. Moreover, choosing a different potential will result in $\beta_{\text{dim-6}}\neq\beta_{\text{dim-4}}$, opening the possibility for interesting physics from $\beta_{\text{dim-6}}$. However, we still expect this operator to be highly suppressed.

As we see in Fig. \ref{fig:Profiles}, rotation angle shapes differ from the universal cases, depending on the 3-form field configuration in the early universe. Interestingly, the large-field case is 
reminiscent of ALP dark matter, in that the birefringence is effectively zero  at low redshift~\cite{Nakatsuka:2022epj}. 
Thus, 3-form dark energy naturally spans the two observational universality classes of the birefringence profile reachable by CMB as well as radio galaxies usually associated with ALP dark energy and ALP dark matter birefringence, with the interpolation controlled by the initial field configuration rather than the particle mass. In our 3-form case, a coupling to dark energy is what produces the birefringence, regardless of the shape of $\beta(z)$. The shape instead acts as a probe of the field configuration in the early universe, which otherwise could be obscured, as each of the three trajectories we considered can effectively mimic $\Lambda$CDM.


\textit{Acknowledgments}
We thank Heliudson Bernardo, Antonio De Felice, Eiichiro Komatsu, Fumihiro Naokawa, and Roman Pasechnik for useful discussions while this work was in progress. The work of K.K. was supported by the National Natural Science Foundation of China (NSFC) under Grant No.~W2532007 and 12547104, and by JSPS KAKENHI Grant-in-Aid for Challenging Research (Exploratory) JP23K17687.

\break 

\bibliographystyle{hunsrt}
\bibliography{refs}

@article{Matsuoka:2026qya,
    author = "Matsuoka, Tomoki and Nomura, Kimihiro and Omiya, Hidetoshi",
    title = "{Reflection polarization of close binaries as a probe of axion dark matter birefringence}",
    eprint = "2607.04550",
    archivePrefix = "arXiv",
    primaryClass = "astro-ph.CO",
    month = "7",
    year = "2026"
}

@article{PhysRevLett.62.82,
  title = {Effective-Field-Theory Model for the Fractional Quantum Hall Effect},
  author = {Zhang, S. C. and Hansson, T. H. and Kivelson, S.},
  journal = {Phys. Rev. Lett.},
  volume = {62},
  issue = {1},
  pages = {82--85},
  numpages = {0},
  year = {1989},
  month = {Jan},
  publisher = {American Physical Society},
  doi = {10.1103/PhysRevLett.62.82},
  url = {https://link.aps.org/doi/10.1103/PhysRevLett.62.82}
}

@article{Green:1996bh,
    author = "Green, Michael B. and Hull, Christopher M. and Townsend, Paul K.",
    title = "{D-brane Wess-Zumino actions, t duality and the cosmological constant}",
    eprint = "hep-th/9604119",
    archivePrefix = "arXiv",
    reportNumber = "RU-96-20, DAMTP-96-41, QMW-PH-96-6",
    doi = "10.1016/0370-2693(96)00643-0",
    journal = "Phys. Lett. B",
    volume = "382",
    pages = "65--72",
    year = "1996"
}

@article{Blasi:1999hu,
    author = "Blasi, Pasquale and Burles, Scott and Olinto, Angela V.",
    title = "{Cosmological magnetic fields limits in an inhomogeneous universe}",
    eprint = "astro-ph/9812487",
    archivePrefix = "arXiv",
    doi = "10.1086/311958",
    journal = "Astrophys. J. Lett.",
    volume = "514",
    pages = "L79--L82",
    year = "1999"
}

@article{Carilli:2001hj,
    author = "Carilli, C. L. and Taylor, G. B.",
    title = "{Cluster magnetic fields}",
    eprint = "astro-ph/0110655",
    archivePrefix = "arXiv",
    doi = "10.1146/annurev.astro.40.060401.093852",
    journal = "Ann. Rev. Astron. Astrophys.",
    volume = "40",
    pages = "319--348",
    year = "2002"
}

@article{Kosowsky:1996yc,
    author = "Kosowsky, Arthur and Loeb, Abraham",
    title = "{Faraday rotation of microwave background polarization by a primordial magnetic field}",
    eprint = "astro-ph/9601055",
    archivePrefix = "arXiv",
    reportNumber = "CFA-4273",
    doi = "10.1086/177751",
    journal = "Astrophys. J.",
    volume = "469",
    pages = "1--6",
    year = "1996"
}

@article{Finelli:2008jv,
    author = "Finelli, Fabio and Galaverni, Matteo",
    title = "{Rotation of Linear Polarization Plane and Circular Polarization from Cosmological Pseudo-Scalar Fields}",
    eprint = "0802.4210",
    archivePrefix = "arXiv",
    primaryClass = "astro-ph",
    doi = "10.1103/PhysRevD.79.063002",
    journal = "Phys. Rev. D",
    volume = "79",
    pages = "063002",
    year = "2009"
}

@article{Harari:1992ea,
    author = "Harari, Diego and Sikivie, Pierre",
    title = "{Effects of a Nambu-Goldstone boson on the polarization of radio galaxies and the cosmic microwave background}",
    reportNumber = "UFIFT-HEP-92-9",
    doi = "10.1016/0370-2693(92)91363-E",
    journal = "Phys. Lett. B",
    volume = "289",
    pages = "67--72",
    year = "1992"
}

@article{Clark:2021kze,
    author = "Clark, S. E. and Kim, Chang-Goo and Hill, J. Colin and Hensley, Brandon S.",
    title = "{The Origin of Parity Violation in Polarized Dust Emission and Implications for Cosmic Birefringence}",
    eprint = "2105.00120",
    archivePrefix = "arXiv",
    primaryClass = "astro-ph.GA",
    doi = "10.3847/1538-4357/ac0e35",
    journal = "Astrophys. J.",
    volume = "919",
    number = "1",
    pages = "53",
    year = "2021"
}

@article{Keating_2012,
   title={SELF-CALIBRATION OF COSMIC MICROWAVE BACKGROUND POLARIZATION EXPERIMENTS},
   volume={762},
   ISSN={2041-8213},
   url={http://dx.doi.org/10.1088/2041-8205/762/2/L23},
   DOI={10.1088/2041-8205/762/2/l23},
   number={2},
   journal={The Astrophysical Journal},
   publisher={American Astronomical Society},
   author={Keating, Brian G. and Shimon, Meir and Yadav, Amit P. S.},
   year={2012},
   month=Dec, pages={L23} }

@article{Planck:2018vyg,
    author = "Aghanim, N. and others",
    collaboration = "Planck",
    title = "{Planck 2018 results. VI. Cosmological parameters}",
    eprint = "1807.06209",
    archivePrefix = "arXiv",
    primaryClass = "astro-ph.CO",
    doi = "10.1051/0004-6361/201833910",
    journal = "Astron. Astrophys.",
    volume = "641",
    pages = "A6",
    year = "2020",
    note = "[Erratum: Astron.Astrophys. 652, C4 (2021)]"
}

@article{Ruegg:2003ps,
    author = "Ruegg, Henri and Ruiz-Altaba, Marti",
    title = "{The Stueckelberg field}",
    eprint = "hep-th/0304245",
    archivePrefix = "arXiv",
    reportNumber = "UGVA-DPT-2003-04-1106, UGVA-DPT-04-1106",
    doi = "10.1142/S0217751X04019755",
    journal = "Int. J. Mod. Phys. A",
    volume = "19",
    pages = "3265--3348",
    year = "2004"
}

@article{Alexander:2024nvi,
    author = "Alexander, Stephon and Manton, Tucker and McDonough, Evan",
    title = "{Field theory axiverse}",
    eprint = "2404.11642",
    archivePrefix = "arXiv",
    primaryClass = "hep-ph",
    doi = "10.1103/PhysRevD.109.116019",
    journal = "Phys. Rev. D",
    volume = "109",
    number = "11",
    pages = "116019",
    year = "2024"
}

@article{Alexander:2023wgk,
    author = "Alexander, Stephon and Gilmer, Humberto and Manton, Tucker and McDonough, Evan",
    title = "{{\ensuremath{\pi}}-axion and {\ensuremath{\pi}}-axiverse of dark QCD}",
    eprint = "2304.11176",
    archivePrefix = "arXiv",
    primaryClass = "hep-ph",
    doi = "10.1103/PhysRevD.108.123014",
    journal = "Phys. Rev. D",
    volume = "108",
    number = "12",
    pages = "123014",
    year = "2023"
}

@article{Kim:1984pt,
    author = "Kim, Jihn E.",
    title = "{A COMPOSITE INVISIBLE AXION}",
    reportNumber = "SNUHE-84-02",
    doi = "10.1103/PhysRevD.31.1733",
    journal = "Phys. Rev. D",
    volume = "31",
    pages = "1733",
    year = "1985"
}

@article{Marsh:2015xka,
    author = "Marsh, David J. E.",
    title = "{Axion Cosmology}",
    eprint = "1510.07633",
    archivePrefix = "arXiv",
    primaryClass = "astro-ph.CO",
    reportNumber = "KCL-PH-TH-2015-50",
    doi = "10.1016/j.physrep.2016.06.005",
    journal = "Phys. Rept.",
    volume = "643",
    pages = "1--79",
    year = "2016"
}

@article{Arvanitaki:2009fg,
    author = "Arvanitaki, Asimina and Dimopoulos, Savas and Dubovsky, Sergei and Kaloper, Nemanja and March-Russell, John",
    title = "{String Axiverse}",
    eprint = "0905.4720",
    archivePrefix = "arXiv",
    primaryClass = "hep-th",
    doi = "10.1103/PhysRevD.81.123530",
    journal = "Phys. Rev. D",
    volume = "81",
    pages = "123530",
    year = "2010"
}

@article{Svrcek:2006yi,
    author = "Svrcek, Peter and Witten, Edward",
    title = "{Axions In String Theory}",
    eprint = "hep-th/0605206",
    archivePrefix = "arXiv",
    reportNumber = "SLAC-PUB-11894",
    doi = "10.1088/1126-6708/2006/06/051",
    journal = "JHEP",
    volume = "06",
    pages = "051",
    year = "2006"
}

@article{Preskill:1982cy,
    author = "Preskill, John and Wise, Mark B. and Wilczek, Frank",
    editor = "Srednicki, M. A.",
    title = "{Cosmology of the Invisible Axion}",
    reportNumber = "HUTP-82-A048, NSF-ITP-82-103",
    doi = "10.1016/0370-2693(83)90637-8",
    journal = "Phys. Lett. B",
    volume = "120",
    pages = "127--132",
    year = "1983"
}

@article{Weinberg:1977ma,
    author = "Weinberg, Steven",
    title = "{A New Light Boson?}",
    reportNumber = "HUTP-77/A074",
    doi = "10.1103/PhysRevLett.40.223",
    journal = "Phys. Rev. Lett.",
    volume = "40",
    pages = "223--226",
    year = "1978"
}

@article{Wilczek:1977pj,
    author = "Wilczek, Frank",
    title = "{Problem of Strong  $P$  and  $T$  Invariance in the Presence of Instantons}",
    reportNumber = "Print-77-0939 (COLUMBIA)",
    doi = "10.1103/PhysRevLett.40.279",
    journal = "Phys. Rev. Lett.",
    volume = "40",
    pages = "279--282",
    year = "1978"
}

@article{Peccei:1977hh,
    author = "Peccei, R. D. and Quinn, Helen R.",
    title = "{CP Conservation in the Presence of Instantons}",
    reportNumber = "ITP-568-STANFORD",
    doi = "10.1103/PhysRevLett.38.1440",
    journal = "Phys. Rev. Lett.",
    volume = "38",
    pages = "1440--1443",
    year = "1977"
}

@article{Kostelecky:2007zz,
    author = "Kostelecky, V. Alan and Mewes, Matthew",
    title = "{Lorentz-violating electrodynamics and the cosmic microwave background}",
    eprint = "astro-ph/0702379",
    archivePrefix = "arXiv",
    reportNumber = "IUHET-504",
    doi = "10.1103/PhysRevLett.99.011601",
    journal = "Phys. Rev. Lett.",
    volume = "99",
    pages = "011601",
    year = "2007"
}

@article{Kaloper:2026ygk,
    author = "Kaloper, Nemanja",
    title = "{CMB Birefringence from Vacuum Interfaces}",
    eprint = "2605.11065",
    archivePrefix = "arXiv",
    primaryClass = "hep-th",
    month = "5",
    year = "2026"
}

@article{eBOSS:2020yzd,
    author = "Alam, Shadab and others",
    collaboration = "eBOSS",
    title = "{Completed SDSS-IV extended Baryon Oscillation Spectroscopic Survey: Cosmological implications from two decades of spectroscopic surveys at the Apache Point Observatory}",
    eprint = "2007.08991",
    archivePrefix = "arXiv",
    primaryClass = "astro-ph.CO",
    doi = "10.1103/PhysRevD.103.083533",
    journal = "Phys. Rev. D",
    volume = "103",
    number = "8",
    pages = "083533",
    year = "2021"
}

@article{DESI:2024lzq,
    author = "Adame, A. G. and others",
    collaboration = "DESI",
    title = "{DESI 2024 IV: Baryon Acoustic Oscillations from the Lyman alpha forest}",
    eprint = "2404.03001",
    archivePrefix = "arXiv",
    primaryClass = "astro-ph.CO",
    reportNumber = "FERMILAB-PUB-24-0147-PPD",
    doi = "10.1088/1475-7516/2025/01/124",
    journal = "JCAP",
    volume = "01",
    pages = "124",
    year = "2025"
}

@article{DES:2026jmi,
    author = "Abbott, T. M. C. and others",
    collaboration = "DES",
    title = "{Constraints on Dynamical Dark Energy from Multiple Probes in the Full Dark Energy Survey}",
    eprint = "2605.27221",
    archivePrefix = "arXiv",
    primaryClass = "astro-ph.CO",
    reportNumber = "DES-2026-0979, FERMILAB-PUB-26-0306-PPD",
    month = "5",
    year = "2026"
}

@article{Bouhmadi-Lopez:2016dzw,
    author = "Bouhmadi-L{\'o}pez, Mariam and Marto, Jo{\~a}o and Morais, Jo{\~a}o and Silva, C{\'e}sar M.",
    title = "{Cosmic infinity: A dynamical system approach}",
    eprint = "1611.03100",
    archivePrefix = "arXiv",
    primaryClass = "gr-qc",
    doi = "10.1088/1475-7516/2017/03/042",
    journal = "JCAP",
    volume = "03",
    pages = "042",
    year = "2017"
}

@article{Ho:2010aq,
    author = "Ho, Shih-Hao and Kao, W. F. and Bamba, Kazuharu and Geng, C. Q.",
    title = "{Cosmological birefringence due to CPT-even Chern-Simons-like term with Kalb-Ramond and scalar fields}",
    eprint = "1008.0486",
    archivePrefix = "arXiv",
    primaryClass = "hep-ph",
    doi = "10.1140/epjc/s10052-015-3426-5",
    journal = "Eur. Phys. J. C",
    volume = "75",
    number = "5",
    pages = "192",
    year = "2015"
}

@article{DESI:2025zgx,
    author = "Abdul Karim, M. and others",
    collaboration = "DESI",
    title = "{DESI DR2 results. II. Measurements of baryon acoustic oscillations and cosmological constraints}",
    eprint = "2503.14738",
    archivePrefix = "arXiv",
    primaryClass = "astro-ph.CO",
    reportNumber = "FERMILAB-PUB-25-0169-PPD",
    doi = "10.1103/tr6y-kpc6",
    journal = "Phys. Rev. D",
    volume = "112",
    number = "8",
    pages = "083515",
    year = "2025"
}

@article{Yin:2024fez,
    author = "Yin, Weichen Winston and Dai, Liang and Huang, Junwu and Ji, Lingyuan and Ferraro, Simone",
    title = "{New Probe of Cosmic Birefringence Using Galaxy Polarization and Shapes}",
    eprint = "2402.18568",
    archivePrefix = "arXiv",
    primaryClass = "astro-ph.CO",
    doi = "10.1103/PhysRevLett.134.161001",
    journal = "Phys. Rev. Lett.",
    volume = "134",
    number = "16",
    pages = "161001",
    year = "2025"
}

@article{Nakai:2023zdr,
    author = "Nakai, Yuichiro and Namba, Ryo and Obata, Ippei and Qiu, Yu-Cheng and Saito, Ryo",
    title = "{Can we explain cosmic birefringence without a new light field beyond Standard Model?}",
    eprint = "2310.09152",
    archivePrefix = "arXiv",
    primaryClass = "astro-ph.CO",
    reportNumber = "RIKEN-iTHEMS-Report-23",
    doi = "10.1007/JHEP01(2024)057",
    journal = "JHEP",
    volume = "01",
    pages = "057",
    year = "2024"
}

@article{Eskilt:2022cff,
    author = "Eskilt, Johannes R. and Komatsu, Eiichiro",
    title = "{Improved constraints on cosmic birefringence from the WMAP and Planck cosmic microwave background polarization data}",
    eprint = "2205.13962",
    archivePrefix = "arXiv",
    primaryClass = "astro-ph.CO",
    doi = "10.1103/PhysRevD.106.063503",
    journal = "Phys. Rev. D",
    volume = "106",
    number = "6",
    pages = "063503",
    year = "2022"
}

@article{Nakatsuka:2022epj,
    author = "Nakatsuka, Hiromasa and Namikawa, Toshiya and Komatsu, Eiichiro",
    title = "{Is cosmic birefringence due to dark energy or dark matter? A tomographic approach}",
    eprint = "2203.08560",
    archivePrefix = "arXiv",
    primaryClass = "astro-ph.CO",
    doi = "10.1103/PhysRevD.105.123509",
    journal = "Phys. Rev. D",
    volume = "105",
    number = "12",
    pages = "123509",
    year = "2022"
}

@article{Komatsu:2022nvu,
    author = "Komatsu, Eiichiro",
    title = "{New physics from the polarized light of the cosmic microwave background}",
    eprint = "2202.13919",
    archivePrefix = "arXiv",
    primaryClass = "astro-ph.CO",
    doi = "10.1038/s42254-022-00452-4",
    journal = "Nature Rev. Phys.",
    volume = "4",
    number = "7",
    pages = "452--469",
    year = "2022"
}

@article{Eskilt:2022wav,
    author = "Eskilt, J. R.",
    title = "{Frequency-dependent constraints on cosmic birefringence from the LFI and HFI Planck Data Release 4}",
    eprint = "2201.13347",
    archivePrefix = "arXiv",
    primaryClass = "astro-ph.CO",
    doi = "10.1051/0004-6361/202243269",
    journal = "Astron. Astrophys.",
    volume = "662",
    pages = "A10",
    year = "2022"
}

@article{Sherwin:2021vgb,
    author = "Sherwin, Blake D. and Namikawa, Toshiya",
    title = "{Cosmic birefringence tomography and calibration independence with reionization signals in the CMB}",
    eprint = "2108.09287",
    archivePrefix = "arXiv",
    primaryClass = "astro-ph.CO",
    doi = "10.1093/mnras/stac3146",
    journal = "Mon. Not. Roy. Astron. Soc.",
    volume = "520",
    number = "3",
    pages = "3298--3304",
    year = "2023"
}

@article{Lue:1998mq,
    author = "Lue, Arthur and Wang, Li-Min and Kamionkowski, Marc",
    title = "{Cosmological signature of new parity violating interactions}",
    eprint = "astro-ph/9812088",
    archivePrefix = "arXiv",
    reportNumber = "CU-TP-926, CAL-675",
    doi = "10.1103/PhysRevLett.83.1506",
    journal = "Phys. Rev. Lett.",
    volume = "83",
    pages = "1506--1509",
    year = "1999"
}

@article{Nodland:1997cc,
    author = "Nodland, Borge and Ralston, John P.",
    title = "{Indication of anisotropy in electromagnetic propagation over cosmological distances}",
    eprint = "astro-ph/9704196",
    archivePrefix = "arXiv",
    doi = "10.1103/PhysRevLett.78.3043",
    journal = "Phys. Rev. Lett.",
    volume = "78",
    pages = "3043--3046",
    year = "1997"
}

@article{Carroll:1997tc,
    author = "Carroll, Sean M. and Field, George B.",
    title = "{Is there evidence for cosmic anisotropy in the polarization of distant radio sources?}",
    eprint = "astro-ph/9704263",
    archivePrefix = "arXiv",
    reportNumber = "NSF-ITP-97-036",
    doi = "10.1103/PhysRevLett.79.2394",
    journal = "Phys. Rev. Lett.",
    volume = "79",
    pages = "2394--2397",
    year = "1997"
}

@article{Carroll:1989vb,
    author = "Carroll, Sean M. and Field, George B. and Jackiw, Roman",
    title = "{Limits on a Lorentz and Parity Violating Modification of Electrodynamics}",
    reportNumber = "MIT-CTP-1782",
    doi = "10.1103/PhysRevD.41.1231",
    journal = "Phys. Rev. D",
    volume = "41",
    pages = "1231",
    year = "1990"
}

@article{Geng:2007va,
    author = "Geng, C. Q. and Ho, S. H. and Ng, J. N.",
    title = "{Neutrino number asymmetry and cosmological birefringence}",
    eprint = "0706.0080",
    archivePrefix = "arXiv",
    primaryClass = "astro-ph",
    doi = "10.1088/1475-7516/2007/09/010",
    journal = "JCAP",
    volume = "09",
    pages = "010",
    year = "2007"
}

@article{Carralot:2026kps,
    author = "Carralot, Florie and Diego-Palazuelos, Patricia and Duivenvoorden, Adriaan J. and Komatsu, Eiichiro and Krachmalnicoff, Nicoletta and Baccigalupi, Carlo",
    title = "{Is cosmic birefringence due to dark energy or dark matter? Simulation-based inference}",
    eprint = "2602.12019",
    archivePrefix = "arXiv",
    primaryClass = "astro-ph.CO",
    month = "2",
    year = "2026"
}

@article{Shao:2025jhp,
    author = "Shao, Jialiang and Obata, Ippei and Zhang, Dongdong",
    title = "{Implications of Cosmic Birefringence for Multi-Field ALP Dark Matter}",
    eprint = "2512.21888",
    archivePrefix = "arXiv",
    primaryClass = "astro-ph.CO",
    month = "12",
    year = "2025"
}

@article{Diego-Palazuelos:2025dmh,
    author = "Diego-Palazuelos, P. and Komatsu, E.",
    title = "{Cosmic Birefringence from the Atacama Cosmology Telescope Data Release 6}",
    eprint = "2509.13654",
    archivePrefix = "arXiv",
    primaryClass = "astro-ph.CO",
    month = "9",
    year = "2025"
}

@article{Remazeilles:2025wzd,
    author = "Remazeilles, Mathieu",
    title = "{Field-level constraints on cosmic birefringence from hybrid ILC maps combining E- and B-mode channels}",
    eprint = "2507.22109",
    archivePrefix = "arXiv",
    primaryClass = "astro-ph.CO",
    doi = "10.1088/1475-7516/2025/12/013",
    journal = "JCAP",
    volume = "12",
    pages = "013",
    year = "2025"
}

@article{Naokawa:2025shr,
    author = "Naokawa, Fumihiro",
    title = "{Universal Profile for Cosmic Birefringence Tomography with Radio Galaxies}",
    eprint = "2504.06709",
    archivePrefix = "arXiv",
    primaryClass = "astro-ph.CO",
    reportNumber = "RESCEU-6/25",
    doi = "10.1103/srfg-9fdy",
    journal = "Phys. Rev. Lett.",
    volume = "136",
    number = "4",
    pages = "041004",
    year = "2026"
}

@article{ParticleDataGroup:2024cfk,
    author = "Navas, S. and others",
    collaboration = "Particle Data Group",
    title = "{Review of particle physics}",
    doi = "10.1103/PhysRevD.110.030001",
    journal = "Phys. Rev. D",
    volume = "110",
    number = "3",
    pages = "030001",
    year = "2024"
}

@article{Zhou:2023aqz,
    author = "Zhou, Ren-Peng and Huang, Da and Geng, Chao-Qiang",
    title = "{Cosmic birefringence from neutrino and dark matter asymmetries}",
    eprint = "2302.11140",
    archivePrefix = "arXiv",
    primaryClass = "astro-ph.CO",
    doi = "10.1088/1475-7516/2023/07/053",
    journal = "JCAP",
    volume = "07",
    pages = "053",
    year = "2023"
}

@article{Heisenberg:2025jut,
    author = "Heisenberg, Lavinia",
    title = "{A Re-Examination Of Foundational Elements Of Cosmology}",
    eprint = "2512.16934",
    archivePrefix = "arXiv",
    primaryClass = "physics.gen-ph",
    month = "12",
    year = "2025"
}

@article{Koivisto:2009ew,
    author = "Koivisto, Tomi S. and Nunes, Nelson J.",
    title = "{Three-form cosmology}",
    eprint = "0907.3883",
    archivePrefix = "arXiv",
    primaryClass = "astro-ph.CO",
    doi = "10.1016/j.physletb.2010.01.051",
    journal = "Phys. Lett. B",
    volume = "685",
    pages = "105--109",
    year = "2010"
}

@article{Koivisto:2009fb,
    author = "Koivisto, Tomi S. and Nunes, Nelson J.",
    title = "{Inflation and dark energy from three-forms}",
    eprint = "0908.0920",
    archivePrefix = "arXiv",
    primaryClass = "astro-ph.CO",
    doi = "10.1103/PhysRevD.80.103509",
    journal = "Phys. Rev. D",
    volume = "80",
    pages = "103509",
    year = "2009"
}

@article{Kobayashi:2009hj,
    author = "Kobayashi, Tsutomu and Yokoyama, Shuichiro",
    title = "{Gravitational waves from p-form inflation}",
    eprint = "0903.2769",
    archivePrefix = "arXiv",
    primaryClass = "astro-ph.CO",
    reportNumber = "WU-AP-297-09",
    doi = "10.1088/1475-7516/2009/05/004",
    journal = "JCAP",
    volume = "05",
    pages = "004",
    year = "2009"
}

@article{Bouhmadi-Lopez:2025lzm,
    author = "Bouhmadi-L{\'o}pez, Mariam and Chiang, Hsu-Wen and Boiza, Carlos G. and Chen, Pisin",
    title = "{Observational constraints on 3-forms dark energy}",
    eprint = "2512.09991",
    archivePrefix = "arXiv",
    primaryClass = "astro-ph.CO",
    month = "12",
    year = "2025"
}

@article{Bouhmadi-Lopez:2026ytj,
    author = "Bouhmadi-L{\'o}pez, Mariam and Chiang, Hsu-Wen and Boiza, Carlos G. and del R{\'\i}o, Javier Ortega and Broadhurst, Thomas J. and Chen, Pisin",
    title = "{Three-form dark energy: constraints and multi-probe comparison with $\Lambda$CDM}",
    eprint = "2606.27436",
    archivePrefix = "arXiv",
    primaryClass = "astro-ph.CO",
    month = "6",
    year = "2026"
}

@article{DeFelice:2012jt,
    author = "De Felice, Antonio and Karwan, Khamphee and Wongjun, Pitayuth",
    title = "{Stability of the 3-form field during inflation}",
    eprint = "1202.0896",
    archivePrefix = "arXiv",
    primaryClass = "hep-ph",
    doi = "10.1103/PhysRevD.85.123545",
    journal = "Phys. Rev. D",
    volume = "85",
    pages = "123545",
    year = "2012"
}

@article{Morais:2016bev,
    author = "Morais, Jo{\~a}o and Bouhmadi-L{\'o}pez, Mariam and Sravan Kumar, K. and Marto, Jo{\~a}o and Tavakoli, Yaser",
    title = "{Interacting 3-form dark energy models: distinguishing interactions and avoiding the Little Sibling of the Big Rip}",
    eprint = "1608.01679",
    archivePrefix = "arXiv",
    primaryClass = "gr-qc",
    doi = "10.1016/j.dark.2016.11.002",
    journal = "Phys. Dark Univ.",
    volume = "15",
    pages = "7--30",
    year = "2017"
}

\end{document}